\documentclass[12pt,twoside]{article}
\usepackage[T1]{fontenc}
\usepackage[latin1]{inputenc}
\usepackage{times}
\usepackage{graphics}
\usepackage{a4wide}

\title{Pluto: A Monte Carlo Simulation Tool for Hadronic Physics}

\author{I.~Fr\"ohlich$^{1,}$\thanks{Presentation given at the XI International Workshop on Advanced Computing and Analysis Techniques in Physics Research,
   April 23-27 2007,
   Amsterdam, the Netherlands.}~, 
  L.~Cazon~Boado$^{2,}$\thanks{Participant in the GSI International Summer Student
    Program.}~,
  T.~Galatyuk$^{1,2}$,
  V. Hejny$^{3}$, \\
  R.~Holzmann$^2$,
  M.~Kagarlis$^{2,}$\thanks{now at: Ventana Systems, Inc., Harvard.}~,
  W.~K\"uhn$^4$, \\
  J.G.~Messchendorp$^{4,}$\thanks{now at:  Kernfysisch Versneller Instituut, Groningen}~,
  V.~Metag$^4$,
  M.-A.~Pleier$^{4,}$\thanks{now at:  Physikalisches Institut, Universit\"at Bonn.}~,
  W.~Przygoda$^5$,\\
  B.~Ramstein$^6$,
  J.~Ritman$^{4,}$\thanks{now at: Institut f\"{u}r Kernphysik.}~,
  P.~Salabura$^{2,5}$, \\
  J.~Stroth$^{1,2}$ 
  and M.~Sudol$^{1,2}$}
      
\date{}

\newcommand{\classref}[1]{{\texttt{#1}}}

\begin{document}

\maketitle

{$^1$}Institut f\"{u}r Kernphysik,  Johann Wolfgang
  Goethe-Universit\"{a}t, \  60486 Frankfurt, Germany.\\
  {$^2$}Gesellschaft f\"{u}r Schwerionenforschung mbH,   64291 Darmstadt, 
  Germany.\\  
  {$^3$}Institut f\"{u}r Kernphysik, 
    Forschungszentrum J\"ulich, 52428 J\"ulich, Germany.\\
  {$^4$}II.Physikalisches Institut,  Justus Liebig Universit\"{a}t 
  Giessen,   35392 Giessen, Germany.\\
  {$^5$}Smoluchowski Institute of Physics, Jagiellonian University 
  of Cracow,   30059 Cracow, Poland.\\
  {$^6$}Institut de Physique Nucl\'{e}aire d'Orsay, CNRS/IN2P3,  91406 
  Orsay Cedex, France.

\begin{abstract}
{Pluto is a Monte-Carlo event generator designed for hadronic
  interactions from Pion production threshold to intermediate energies of a
  few GeV per nucleon, as well as for studies of heavy ion reactions. The
  package is entirely based on ROOT, without the need of additional packages,
  and uses the embedded C++ interpreter of ROOT to control the event
  production.  The generation of events based on a single reaction chain and
  the storage of the resulting particle objects can be done with a few lines
  of a ROOT-macro. However, the complete control of the package can be taken
  over by the steering macro and user-defined models may be added without a
  recompilation of the framework. Multi-reaction cocktails can be facilitated as
  well using either mass-dependent or user-defined static branching ratios.
      
   The included physics uses resonance production with mass-dependent
   Breit-Wigner sampling. The calculation of partial and total widths for
   resonances producing unstable particles is performed recursively in a
   coupled-channel approach. Here, particular attention is paid to the
   electromagnetic decays, motivated by the physics program of HADES. The
   thermal model supports 2-component thermal distributions, longitudinal
   broadening, radial blast, direct and elliptic flow, and impact-parameter
   sampled multiplicities.  
   
   The interface allows angular distribution models (e.g. for the primary
   meson emission) to be attached by the user as well as descriptions of
   multi-particle correlations using decay chain templates.  The exchange of
   mass sampling or momentum generation models is also possible.  For elementary reactions,
   angular distribution models for selected channels are already part of the
   framework, based on parameterizations of existing data.

   This report gives an overview of the design of the package, the included
   models and the user interface.}

\end{abstract}

\section{Preface}

 Simulations are an integral part of experimental programs associated with
 scattering experiments and particle accelerators. Such studies are required
 both in order to understand the properties of experimental setups (e.g.
 reaction-dependent acceptances of various types of detectors), as well as to
 gain insight into the processes of interest, so that relevant experiments may
 be optimized and experimental spectra may be interpreted.
 
 In a first step of such a simulation the primary reaction has to be described,
 which means that a number of particle tracks (Lorentz vectors) are generated.
 These tracks form the events, which are further processed through a
 digitization package in order to get the secondary particles coming from the
 interactions of the particles with the detector material. In such a way, the
 response of a detector system for a certain event topology can be studied. By
 adding realistic background, the sensitivity of a detector array on a
 specified process can be verified, which is normally done before
 carrying out the experimental run.  Several different models have to be included
 in the event generator, either to describe the reactions which are under study, or a
 number of background channels.
 
 But event generators are also very useful after data have been taken and
 should be interpreted.  Here, the caveat is that the degrees of freedom of a
 reaction are correlated. Moreover, the efficiency and acceptance of
 experimental setups are functions of all phase space parameters. Due to
 energy and momentum conservation, this has in particular a large effect in
 exclusive reactions. By projecting the multi-dimensional result of the
 experiment to the observable quantities under study, and correcting this
 result for acceptance without a detailed knowledge of the detector
 performance and the included physics of the hidden degrees of freedom, the
 resulting interpretation might be artificially changed.
 
 This immediately leads to the conclusion that the observable quantities which
 are already well known have to be included into event generators in order
 to address the remaining open questions.  Therefore, each event generator
 should be adapted to the physics program for the corresponding experiments.

 A particular challenge for the generation of events is the theory of the
 Quantum Chromo Dynamics (QCD), since its coupling constant $\alpha_s$ 
 strongly depends on the momentum transfer: In high momentum transfer experiments
 partonic degrees of freedom can be studied, whereas in low momentum transfer experiments
 the hadrons itself can be treated as effective particles.
 
 For certain fields in the QCD, specified event generators are already
 existing. We will not go through the complete list, a review on high-energy
 event generators, including the description of the hard interactions of
 partons, matrix element generators and hadronization packages can be found
 in~\cite{le_houches}. For the hadronization part, which comes closer to the
 experiments studying hadronic interactions, Pythia~\cite{pythia} is one of
 the frequently used packages. In addition, the EvtGen~\cite{evtgen} package
 was developed for the production of B-mesons, which has a user-friendly decay
 model interface and thus includes a large number of decay processes. However,
 in the context of hadronic interactions at lower energies comprehensive event
 generators are lacking.
 
 The package ``Pluto''~\cite{kagarlis,download} presented in this report is
 geared towards elementary hadronic as well as heavy-ion induced reactions at
 intermediate to moderately high energies, mainly motivated by the physics
 program of the HADES\footnote{``High Acceptance Di-Electron
   Spectrometer''}~\cite{hades_prl} experiment, which is installed at the SIS
 synchrotron of the GSI. As the HADES experiment has published the first data
 and successfully finished various experimental runs, the need for realistic
 and detailed simulations is evident and growing.  Pluto is an available,
 standardized and efficient tool that facilitates such simulations.
 Moreover, it can be adapted and integrated into simulation
 environments for other experiments. In particular it has been used for the
 simulations in the context of the planned CBM\footnote{``Compressed Baryonic
   Matter''} experiment~\cite{senger} which is going to be operated at the new
 FAIR\footnote{``Facility for Antiproton and Ion research''} facility.

 
 Starting from the basic philosophy, that an event generator has to fulfill
 different tasks during the life-time of an experiment, the Pluto framework
 was designed to have a standard user-interface allowing for quick studies,
 but can be changed on the other hand in such a way to include sophisticated
 new models, including coupled channel approaches as well as interferences
 between various channels.
 
 This report is structured in the following way: First, we give an overview of
 the desired calculations for the physics of hadronic interactions at low
 energies. One of the main features, namely the mass-dependent widths of the
 resonances, is described in Sec.~\ref{physics}.  In addition, the mass
 sampling of the virtual photons and their decay into $e^+e^-$ will be
 discussed in Sec.~\ref{dileptons}.  In Sec.~\ref{angular}, the angular
 correlation parameterizations and multi-particle correlations are shown,
 which have been utilized for selected processes. The thermal model, used in
 the HADES heavy-ion program~\cite{hades_prl}, is explained in
 Sec.~\ref{thermal}. In a second part we go more into the technical
 realization of the package and show the user interface for event production,
 starting from some simple examples up to the level of generating cocktails in
 Sec.~\ref{interface}.  Finally, in Sec.~\ref{customization} the internal
 structure of Pluto is described, whose features allow for the extension of
 the included data base and the introduction of user-defined models.

\section{Resonance mass distributions}\label{physics}

An important effect that must be taken into account for realistic simulations
of hadronic interactions at low energies is the deviation of resonance shapes
from fixed-width Breit-Wigner distributions, which is typically modeled as a
mass-dependence in the resonance width. This is particularly important for
resonances with large widths (i.e. predominantly strongly decaying), such as
the $\rho$, $\Delta$, $N^*$ and $\Delta^*$ resonance excitations for which the
effect is largest. The next few subsections discuss the formalism behind the
calculation of partial and total widths done in the Pluto code.

Following the usual Ansatz (see e.g.~\cite{cite_9_teis}) we use the
relativistic form of the Breit Wigner distribution:
\begin{equation}\label{eqn:breit_wigner}
g(m) = A \frac{m^2 \Gamma^{\rm tot}(m)} {(M^2_{\rm R} - m^2)^2 + m^2 (\Gamma^{\rm tot}(m))^2}
\end{equation}
where $m$ denotes the running unstable mass, and $M_{\rm R}$ is the static pole mass
of the resonance. The mass-dependent width depends on the partial widths:
\begin{equation}\label{eqn:width_sum}
\Gamma^{\rm tot}(m) = \sum^N_{\rm k} \Gamma^{\rm k}(m)
\end{equation}
with $N$ the number of decay modes. The factor $A$ has been chosen such that
the integral is statistically normalized ($\int dm\ g(m) = 1$). 

Eqn.~(\ref{eqn:breit_wigner}) is used between a minimum $m_{\rm min}$ and
maximum mass $m_{\rm max}$ which is set for each particle individually in the
data base, thus avoiding to sample masses which are extremely off-shell. This
range is $[M_{\rm R}-2\Gamma^{\rm tot}, M_{\rm R}+12\Gamma^{\rm tot}]$ by
default, but the limits can be changed by the user.

\subsection{Unitarity condition and self-consistent approach}

For those decay modes for which dedicated models are existing in Pluto, the
decay width $\Gamma^{\rm k}(m)$ is calculated explicitly as a function of
mass. Alternatively, they may be added by the user of the package, as
described in Sec.~\ref{customization}. The included Breit-Wigner model is
interfacing to these decay models via a strictly object-oriented design. This
makes sure to have always a self-consistent result.

The known decay modes have an implicit energy threshold, below which their
respective decay widths vanish.  This can be deduced in 2 ways: The lowest
invariant mass $M_{\rm th}^{\rm k}$ needed for each decay mode ``k'' is always
the sum of the stable particle masses in the final state (i.e. after all
decays). This is implemented by a recursive call of all involved nested decays
until stable particles are reached, and taking the lowest available mass sum
as the invariant mass threshold $M_{\rm th}^{\rm k}$, which ensures e.g.  that
calculated mass-dependent partial widths are zero below the threshold.  

For those modes, which have no dedicated model, the fixed static partial width
is used between the mass threshold $M_{\rm th}^{\rm k}$ (or the minimum mass
$m_{min}$ of the decay parent if larger) and the maximum mass $m_{max}$ of the
decay parent.

This leads the the following condition for the mass-dependent  branching
ratio:
\begin{equation}\label{eqn:br}
  b^{\rm k}(m)=\left\{ 
    \begin{array}{cc}
      \frac{\Gamma^{\rm k}(m)}{\Gamma^{\rm tot}(m)} & ;\ \
      m>{\rm max}(M_{\rm th}^{\rm k},m_{min}) \\
      0 & ;\ \
      m<{\rm max}(M_{\rm th}^{\rm k},m_{min})
    \end{array}
  \right.
\end{equation}
In a second step, Eqn.~(\ref{eqn:width_sum}) is used with $m=M_{\rm R}$ to
check the partial widths at the mass pole where in many cases the branching
ratios are defined, and scale each partial width model such that it matches
this condition at the mass pole.

In selected cases it is not useful to define the branching ratio sharply at
the mass pole, but as the total fraction of the integrated yield $\int d m\ 
g^{\rm k}(m)$. This is in particular needed when particle production occurs
at or above the mass pole of the parent. Such an example is the decay mode
$N^*(1535) \to \Delta(1232) \pi$ which has been seen, but the branching ratio is
only known with an upper limit and thus it is impossible to fix the partial
width at the mass pole.  By combining Eqn~\ref{eqn:br} with the normalized
distribution function~\ref{eqn:breit_wigner}, the static branching ratio can
be recalculated as the integral of the weight folded with the mass-dependent
branching ratio:
\begin{equation}\label{eqn:scbr}
  b^{\rm k}=\frac{\Gamma^{\rm k}}{\Gamma^{\rm tot}}
    =\frac{\int d m\ g(m)\cdot \frac{\Gamma^{\rm k}(m)}{\Gamma^{\rm tot}(m)}}{\int
    d m\ g(m)}
\end{equation}
and combined with the total normalization the partial decay width can be rewritten as:
\begin{equation}\label{eqn:scpw}
  \Gamma^{\rm k} = \int dm\ g(m)\cdot \frac{\Gamma^{\rm k}(m)}{\Gamma^{\rm
  tot}(m)} \Gamma^{\rm tot}
\end{equation}
The mass-dependent branching ratios used in the Breit-Wigner
Eqn.~\ref{eqn:breit_wigner} result in asymmetric mass spectra shapes, but
%
%
the Eqns.~(\ref{eqn:scbr}-\ref{eqn:scpw}) give some constraints which are
checked during the initialization.  Since they contain the distribution
functions $g(m)$ which are based originally on Eqn.~(\ref{eqn:width_sum}) and
thus all decay widths, this is done iteratively until all required conditions
(the total normalization as well as the chosen branching ratio definition) are
fulfilled or a break condition has been reached, which serves as a
self-consistent calculation.


\subsection{Two-body hadronic decays in stable products}

The majority of well-established resonance hadronic decays involve channels
with two decay products, whereas multi-product decay modes might
considered to be the outcome of a series of successive two-body decays through
intermediate resonances~\cite{cite_9_teis,cite_7_wolf,  cite_11_wolf,
  cite_22_pdg}. This is also the approach taken in Pluto, where two-body
hadronic decay widths are calculated explicitly.

This width is derived from a well-known ansatz~\cite{cite_22_pdg}, which has
been established in particular for the
$\Delta(1232)$-resonance~\cite{cite_9_teis, cite_7_wolf, cite_8_monitz}:
\begin{equation}\label{eqn:width_monitz}
  \Gamma^{\rm k}_{m_1 m_2}(m) =  x_{M_{\rm R}}(m)
  \left(\frac{q^{\rm R}_{m_1 m_2}(m)}{q^{\rm R}_{m_1 m_2}(M_{\rm R})} \right)
  ^{2L_{\rm tr}+1}
  \left( \frac{ \nu^{\rm R}_{m_1 m_2}(m)}{\nu^{\rm R}_{m_1 m_2}(M_{\rm R})} \right)
  \Gamma^{\rm k}_{\rm R}
\end{equation}
where the subscript R in general refers to resonance observables corresponding
to the static mass pole for the decay mode in hand, which for the
$\Delta$-resonance e.g. are $M_{\rm R}=$1.232~GeV/c$^2$ and $\Gamma^{\rm
  tot}_{\rm R}=120$~MeV, whereas un-subscripted variables refer to the
corresponding actual-mass observables. The dependence on the two decay
products with the masses $m_1$ and $m_2$ enters via the terms $q^{\rm R}(m)$
and $q^{\rm R}(M_{\rm R})$, namely the (equal in absolute value) momenta of
one out of the two decay products in the rest frame of the parent resonance R.
In the case of two stable products, the masses $m_1$ and $m_2$ are directly
related to the decay mode ``k''.

For the $\Delta(1232)$-resonance, calculating the decay width is a simple
matter since its decay strength is essentially exhausted in the $N\pi$ channel
(branching ratio>99\%).  This means, that $\Gamma^{p\pi}_{m_p m_\pi}$
coincides with the total width.  The pion and nucleon masses $m_p,m_\pi$ may
be considered fixed for all purposes, since the pion has no strong decay modes
and its width has a negligible effect in Eqn.~(\ref{eqn:width_monitz}).  In
general, for arbitrary resonances other than the $\Delta(1232)$-resonance,
$\Gamma^{\rm k}$ refers to the decay width of the parent resonance with mass
equal to the mass pole via the decay mode specified by the identity of the two
decay products and the transition angular momentum.  The dependence on the
decay mode enters via the angular momentum transfer in the $2L_{\rm tr}+1$
exponent of the ratio $\frac{q^{\rm R}_{m_1 m_2}(m)}{q^{\rm R}_{m_1
    m_2}(M_{\rm R})}$ in Eqn.~(\ref{eqn:width_monitz}), with $L_{\rm tr}$ the
transferred orbital angular momentum for the resonance decay. In the case of
the $\Delta$-resonance, which is almost entirely in $p$-wave, $L_{\rm tr}$ is
equal to 1. However, the transfer angular-momentum dependence is in general a
non-trivial matter, since several multi-poles may be interfering. Nonetheless,
data are sparse to non-existent for high multi-pole transitions, and it is
generally a reasonable approximation to treat the decay width as entirely due
to the lowest-allowed multi-pole. This has been done on the basis of
angular-momentum coupling and parity considerations~\cite{cite_10_friman}.

Variants of Eqn.~(\ref{eqn:width_monitz}) are also encountered in the
literature, particularly with regard to the expression in the right-most
bracket that represents the effective cutoff. We follow
Ref.~\cite{cite_9_teis} which uses for the resonance the cutoff
parameterization of Ref.~\cite{cite_8_monitz}, with
\begin{equation}\label{eqn:cutoff_monitz}
  \nu^{\rm R}_{m_1 m_2}(m)=\frac{\beta^2}{\beta^2+(q^{\rm R}_{m_1 m_2}(m))^2}
\end{equation}
and the parameter $\beta$=300~MeV for $\Delta(1232)$ and mesons decays, among
with the phase space factor:
\begin{equation}\label{eqn:m0_over_m}
  x_{M_{\rm R}}(m)=\frac{M_{\rm R}}{m}
\end{equation}
For higher baryon resonances, the parameterization
\begin{equation}\label{eqn:cutoff_res}
  \beta = (M_{\rm R} - m_1 - m_2)^2+\frac{(\Gamma_{\rm R}^{\rm tot})^2}{4},\ \
  x=1
\end{equation}
has been applied.

\subsection{Two-body hadronic decays in unstable products}

The greatest complication, however, which arises in the general case of a
resonance decay with either or both of the decay products unstable, is due to
the fact that the product masses are in general not fixed as in the case of
the $\Delta(1232)$ resonance, but can take values from a distribution function.

In Ref.~\cite{cite_9_teis} and elsewhere few specific cases, with one of the
decay products unstable but decaying to stable products and the other with
fixed mass, are treated explicitly.  In general, calculating the decay width
when both the decay products are unstable hadrons, or with arbitrarily many
embedded decays, is a rather complicated matter. In Pluto, these cases are
treated explicitly, making it possible to calculate realistic spectral
functions for heavy $N^*$ resonances with multiple decay modes and a large
``depth'', i.e.  nested unstable hadron decays.  This is done as follows: Let
us first consider the case $H_p \to h_1 + h_2$ where one of the decay products
(say the first) is an unstable hadron and the other has a fixed mass ($m_2$).
The decay width becomes:
\begin{equation}\label{eqn:m1_width}
\Gamma^{\rm k}_{m_2}(m) = \int_{m_{min}}^{m_{max}}dm_1\ g^{h_1}(m_1)\cdot
q^{H_p}_{m_2}(m,m_1)\cdot \Gamma^{\rm k}_{m_1 m_2}(m)
\end{equation}
where we have introduced the dependence on the product masses explicitly in
$\Gamma^{\rm k}_{m_1 m_2}(m)$.  The product of the daughters mass shape
$g^{h_1}(m_1$), which is the Breit-Wigner Eqn.~\ref{eqn:breit_wigner}, with
the momentum $q^{H_p}_{m_2}(m,m_1)$ of one selected product in the rest frame
of the parent particle introduces the product-mass dependence explicitly. It
works as the effective distribution function for the mass variable $m_1$, and
weighs properly the contribution of each combination of product masses with
the corresponding decay width $\Gamma^{\rm k}_{m_1 m_2}(m)$. The momentum term
acts as a phase space factor and is effectively a cutoff which guarantees a smooth
falloff at the kinematical limit~\cite{cite_11_wolf, cite_10_friman}.

\begin{figure}
\begin{center}
\resizebox{0.59\columnwidth}{!}{%
  \includegraphics{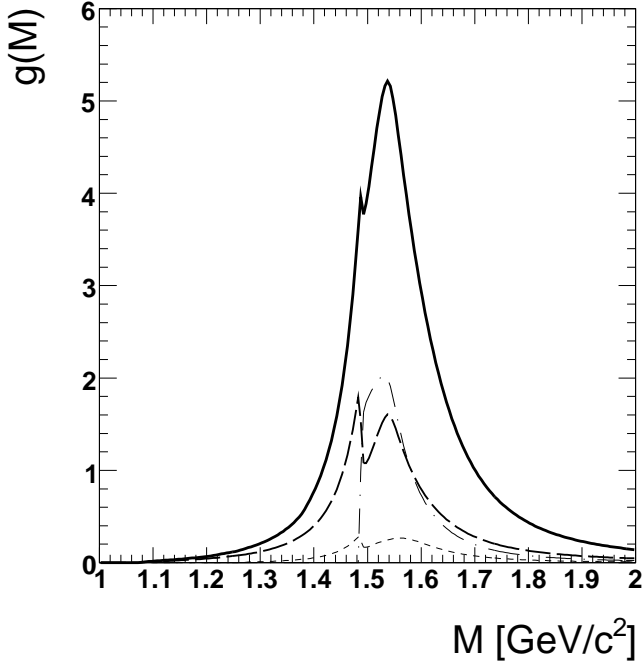}
}
\begin{minipage}[b]{0.40\columnwidth}
\caption{The free
  $N^*(1535)$ total spectral shape (solid line) among with selected partial
  decay shapes as calculated by the Pluto framework: dashed line: $N^*(1535)
  \to N + \pi$, short dashed line: $N^*(1535) \to N^*(1440) + \pi$,
  dotted-dashed line: $N + \eta$.  The first one exhibits a cups-like
  structure near the $\eta$ threshold.\newline ~}
\label{fig:n1535_partial_contribution}       
\end{minipage}
\end{center}
\end{figure}

This expression is generalized in the case of two unstable hadrons:
\begin{equation}\label{eqn:m2_width}
\Gamma^{\rm k}(m) = \int_{m_{min1}}^{m_{max1}}dm_1\ g_1(m_1)
\int_{m_{min2}}^{m_{max2}}dm_2\ g_2(m_2)\cdot
q^{H_p}(m,m_1,m_2) \cdot\Gamma^{\rm k}_{m_1 m_2}(m)\cdot \theta(m>m_1+m_2)
\end{equation}
where the Breit-Wigner shape of the second unstable hadron is accounted for as
well. The Eqns. (\ref{eqn:m1_width}, \ref{eqn:m2_width}) must be normalized,
namely the probability distribution must be integrated over the full range of
validity. The normalization factor is obtained by removing $\Gamma^{\rm
  k}_{m_1 m_2}(m)$ from Eqns.~(\ref{eqn:m1_width}, \ref{eqn:m2_width}) and
evaluating the remaining integral which includes all distribution functions
folded with the 2-body phase space factor $q^{H_p}$.  Subsequently,
corresponding normalization factors are divided out. For an unstable particle
the mass range is defined in such a way that the sampling function $g(m)$ in
the sampling procedure takes values from its mass pole $g_{max}$ down to
$g(m)>0.01 \cdot g_{max}$ which avoids numerical instabilities and shows the
correct behavior in the limit of small widths. The step function
$\theta(m>m_1+m_2)$ is used to fulfill energy conservation.

An example of the calculations described above can be seen in
Fig.~\ref{fig:n1535_partial_contribution}. Here, the $N^*(1535)$ is shown with
selected contributions of partial decays. Obviously, already the existence of
the $N\eta$ decay channel build structures in the $N\pi$
exit channel. Consequently, when simulating the reaction $NN \to N^*(1535) N
\to NN\pi$, this structure should be observed in the $N\pi$ invariant mass.
This is indeed an important feature of the simulation involving broad
resonances: The decay modes cannot be treated independently but should be
rather combined in a coupled-channel calculation.

In the case of three decay products, the static, rather than mass-dependent widths
and branching ratios from the data base are used. 


\subsection{Mass sampling in the hadronic decays}\label{m3_model}

For the decay sampling actually the Breit-Wigner distribution $g(m)$ as given
in Eqn.~(\ref{eqn:breit_wigner}) is convoluted with a phase-space factor for
the sampling of the mass of an unstable hadron $h_1$ in a decay $H_p \to h_1 +
h_2$ where $H_p$ is the parent resonance and $h_2 $ is a stable hadron:
\begin{equation}\label{eqn:m1_mass}
  G^{H_p \to h_1 + h_2}_{m_p,m_2}(m_1) = g^{h_1}(m_1)\cdot q^{H_p}_{m_p,m_2}(m_1)
\end{equation}
assuming that $m_p$ (the mass of the parent resonance) has been obtained
before by a sampling method on its distribution function $g^{H_p}(m_p)$. This
is not a trivial matter: When sampling the parent resonance, in principle the
decay products should be known in order to sample the correct shape as
discussed in the previous subsection. This has the further consequence that in
Eqn.~(\ref{eqn:m1_mass}) the distribution $g^{h_1}(m_1)$ has to be replaced by
$g^{h_1 \to k}(m_1)$ if particle $h_1$ is unstable and will decay via a
consecutive channel ``k'' (just assume e.g.  that $h_1$ is the $N^*(1535)$ but
the decay into $N\pi$ should be considered as shown in
Fig.~\ref{fig:n1535_partial_contribution}).  Consequently $m_p$ must have been
sampled before by using the distribution $g^{H_p \to h_1h_2}(m_p)$ which takes
into account explicitly our chosen decay mode.

\begin{figure}
\begin{center}
\resizebox{0.59\columnwidth}{!}{%
  \includegraphics{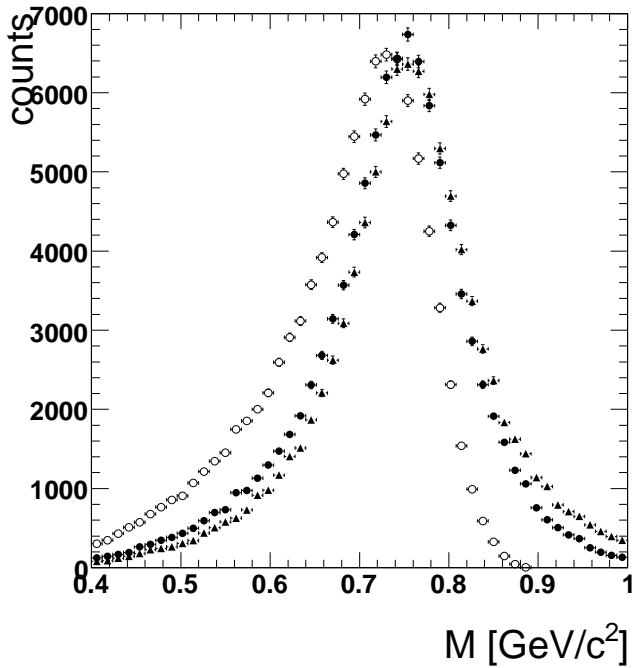}
}
\begin{minipage}[b]{0.40\columnwidth}
  \caption{The $\rho^0$ meson spectrum obtained by a Pluto simulation
    of the $pp \to pp \rho^0$ reaction at three different kinetic beam energies: Open circles:
    2.2 GeV, solid circles: 2.85 GeV and triangles: 3.5 GeV. The influence of
    the three-body phase space is clearly visible.  \newline ~}
\label{fig:rho_sample_phase}       
\end{minipage}
\end{center}
\end{figure}

If the second particle itself is also unstable, the expression is generalized
as:
\begin{equation}\label{eqn:m2_mass}
G^{H_p \to h_1 + h_2}_{m_p}(m_1,m_2) = g^{h_1}(m_1)\cdot q^{H_p}_{m_p}(m_1,m_2)\cdot g^{h_2}(m_2)
\end{equation}
Again, if consecutive decay modes
follow for the particles $h_1,h_2$, the probability distribution functions
$g^{h}(m)$ have to be replaced by $g^{h \to \rm k}(m)$, respectively. This is
ensured automatically by the Pluto framework in all cases where a particular
decay chain has been set up by the user.

For three-body decays, the phase space cannot be calculated only by the parent
mass, therefore a different approach was taken: First all daughter particles
are sampled according to Eqn.~(\ref{eqn:breit_wigner}). Using these masses,
the three-body phase space is calculated, which is the projection of the
Dalitz plane on e.g. $M_{h_1,h_2}^2$. By using the rejection method (the test
function is a constant value with $(M_{h_1,h_2}^{\rm max})^2$) broad resonance
like the $\rho^0$ exhibit the correct shape at the phase space limit (see
Fig.~\ref{fig:rho_sample_phase}).

\section{Production of virtual photons}\label{dileptons}

\subsection{Dalitz decay of pseudoscalar mesons}\label{etadalitz}

The total width, Eqn.~(\ref{eqn:width_sum}), involves the sum of partial decay
widths, only some of which are of the form~(\ref{eqn:width_monitz},
\ref{eqn:m1_width}, \ref{eqn:m2_width}) for decay modes involving two hadrons
as products. Another process of interest for the HADES physics program is the
Dalitz decay of pseudoscalar mesons.  In these processes one of the two decay
products is a virtual (massive) photon $\gamma^*$, which subsequently decays
to a dilepton pair.  These are on one hand sources of the continuum in
dilepton invariant mass spectra~\cite{hades_prl}, but are on the other hand on
their own accord probes of electromagnetic form factors.  The processes of
interest are $\pi^0,\eta,\eta' \to \gamma\gamma^* \to \gamma e^+e^-$.

\begin{figure}
\begin{center}
\resizebox{0.49\columnwidth}{!}{%
  \includegraphics{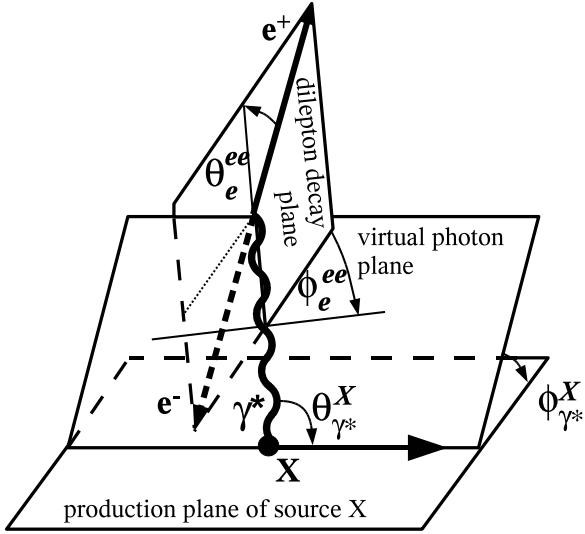}
}
\begin{minipage}[b]{0.50\columnwidth}
\caption{Overview  of the dilepton properties: In addition to the 
  invariant mass $m^{\rm inv}_{\gamma^*}$ and the momentum $p^{X}_{\gamma^*}$,
  4 angles have to be taken into account, depending on the production plane
  and momentum of the source $X$ (Fig. taken from~\cite{ingo}). \newline ~ }
\label{fig:dilepton_angle}       
\end{minipage}
\end{center}
\end{figure}

In this context, it should be pointed out that a virtual photon (decaying into
2 stable particles) has 6 degrees of freedom, which are outlined in
Fig.~\ref{fig:dilepton_angle}: Beside the invariant mass $m^{\rm inv}_{\gamma^*}$
these are the momentum $p^{X}_{\gamma^*}$, the polar $\theta^{X}_{\gamma^*}$
and the azimuthal emitting angle $\phi^{X}_{\gamma^*}$ of the virtual photon
in the rest frame of the source $X$. In addition, the 2 decay angles of the
photon into dilepton pairs, which are usually described with the helicity
angle $\theta^{ee}_e$, and the Treiman-Yang angle $\phi^{ee}_e$.

For the pseudoscalar mesons, which are spin-less, no alignment information can
be carried from the production mechanism to the decay, so
$\theta^{X}_{\gamma^*}$, $\phi^{X}_{\gamma^*}$ and $\phi^{ee}_e$ are
isotropic. The helicity angle distribution however is calculated for
pseudoscalar mesons to be $1+\cos^{2}\theta^{ee}_e$~\cite{brat}, which is
included in Pluto by default.

Since the pseudoscalar mesons have a negligible small width the momentum
$p^{X}_{\gamma^*}$ is in this case fully determined by the invariant mass of
the virtual photon $m^{\rm inv}_{\gamma^*}$, which is also the invariant mass of
the final dilepton pair.

For pseudoscalar-meson Dalitz decays the mass dependence of the Dalitz-decay
width is given by~\cite{cite_6_ernst,cite_12_landsberg}:
\begin{equation}\label{eqn:ps_dalitz_decay}
  \frac{d\Gamma^{\rm k}(m)}{\Gamma^{A\to2\gamma}dm}
  =\frac{4\alpha}{3\pi m}\sqrt{1-\frac{4m_e^2}{m^2}}
  \left(1+\frac{2m_e^2}{m^2} \right)
  \left(1-\frac{m^2}{m_A^2} \right)
  \left|F_A(m^2)\right|^2
\end{equation}
where the index $A$ refers to the (parent) pseudoscalar meson, and $m$,
$m_e$, and $m_A$ are the dilepton, electron, and pseudoscalar masses, and
$F_A(m^2)$ is the parent form factor:

\begin{itemize}

\item For the $\pi^0$:
\begin{equation}\label{eqn:pi0_dalitz_decay}
f_{AB}(m^2) \approx 1+m^2\left.\frac{dF_{AB}}{dm^2} \right|_{m^2 \to 0}
=1+m^2b_{AB}, \ \ b_{AB}=b_{\pi^0}=5.5 \pm 1.6\ {\rm GeV}^{-2}
\end{equation}

\item For the $\eta$:
\begin{equation}\label{eqn:eta_dalitz_decay}
F(m^2)=\left( 1-\frac{m^2}{\Lambda^2_i} \right)^{-1}, \ \ \Lambda_\eta=0.72 \pm 0.09\ {\rm GeV}
\end{equation}

\item For the $\eta\prime$:
\begin{equation}\label{eqn:etap_dalitz_decay}
\left| F(m^2) \right| ^2 =
\frac{\Lambda^2(\Lambda^2+\gamma^2)}{(\Lambda^2-m^2)+\Lambda^2\gamma^2},
\ \ \Lambda_{\eta\prime}=0.76GeV, \ \ \gamma_{\eta\prime}=0.10\ {\rm GeV}
\end{equation}

\end{itemize}

Eqn. (\ref{eqn:ps_dalitz_decay}) is used as the effective distribution
function from which virtual-photon masses are sampled. Its integral also
yields the partial decay width for pseudoscalar Dalitz-decay modes.

\subsection{Dalitz decay of vector mesons}

Currently only the vector-meson Dalitz decay $\omega\to \gamma^*\pi^0\to
e^+e^-\pi^0$ is implemented in the code. The mass dependence of the decay
width is
\begin{equation}\label{eqn:omega_dalitz_decay}
\frac{d\Gamma^{\rm k}(m)}{\Gamma^{A\to B\gamma}dm}
=\frac{2\alpha}{3\pi m}
\sqrt{1-\frac{4m_e^2}{m^2}}
\left(
  \left(
    1+\frac{m^2}{m_A^2-m_B^2}
  \right)^2-
  \left(
    \frac{2m_Am}{m_A^2-m_B^2}
  \right)
\right)^\frac{3}{2}
\left|F_A(m^2)\right|^2
\end{equation}
where the notation is as in Eqn.~(\ref{eqn:ps_dalitz_decay}), and the new
index $B$ refers to the $\pi^0$. The form factor is as in
Eqn.~(\ref{eqn:etap_dalitz_decay}), with $\Lambda_{\omega}$=0.65~GeV, and
$\gamma_{\omega}$=0.04~GeV.

\subsection{Dalitz decay of the $\Delta(1232)$}

\begin{figure}
\begin{center}
\resizebox{0.59\columnwidth}{!}{%
  \includegraphics{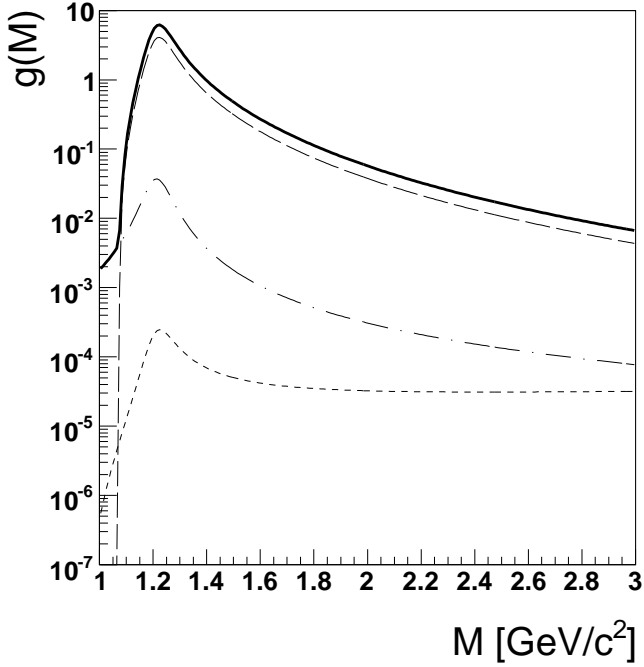}
}
\begin{minipage}[b]{0.40\columnwidth}
  \caption{Free spectral shape $g(m)$ of the $\Delta^+(1232)$ (solid line) as
    a result of the Pluto model calculation, compared to the distribution
    functions for dedicated decay states: Dotted line: $\Delta^+
    (1232)\to e^+e^+p$, dashed line: $\Delta^+ (1232) \to \pi^0+p$ and
    dashed-dotted line: $\Delta^+ (1232) \to \gamma +p$. \newline ~}
\label{fig:delta}       
\end{minipage}
\end{center}
\end{figure}

For $\Delta(1232)\to N\gamma^*\to Ne^+e^-$, the mass-dependence of the width is
calculated directly from the matrix element, without scaling factors as in the
previous two cases~\cite{cite_6_ernst,cite_12_landsberg}:
\begin{equation}\label{eqn:delta_dalitz_decay}
\frac{d\Gamma^{\rm \Delta\to Ne^+e^-}(m)}{dm}
=\frac{2\alpha}{3\pi m}
\sqrt{1-\frac{4m_e^2}{m^2}}
\left(
  1+\frac{m_e^2}{m^2}
\right)
\Gamma^{A\to B\gamma^*}_{m_Am_B}(m)
\end{equation}
where the decay rate is defined as:
\begin{equation}\label{eqn:delta_dalitz_decay2}
  \Gamma^{A\to B\gamma^*}_{m_Am_B}(m)
  =\frac{q^{A}_{m_A m_B}(m)}{8\pi m_A^2}\left|M^{A\to B\gamma^*}\right|^2
\end{equation}
with the indices A and B referring to the parent (resonance) and product
nucleon respectively, and $q^{A}_{m_A m_B}(m)$ the (common) product
center-of-mass momentum in the parent rest frame. The matrix element of
Eqn.~(\ref{eqn:delta_dalitz_decay2}) is~\cite{ernst_bass}:
\begin{equation}\label{eqn:delta_dalitz_decay3}
  \begin{array}{lll}
    \left|M^{A\to B\gamma^*}\right|^2
    & = &
    e^2 G_M^2
    \frac{(m_\Delta+m_N)^2((m_\Delta-m_N)^2-m^2)}
    {4m_N^2((m_\Delta+m_N)^2-m^2)^2} \\ & &
    (7m_\Delta^4 + 14m_\Delta^2m^2 + 3m^4 + \\ & &
    8m_\Delta^3m_N 
    +2m_\Delta^2m_N^2 + 6m^2m_N^2 +3m_N^4)
  \end{array}
\end{equation}
where the index $N$ refers to the produced nucleon, $e$ is the electron charge,
and $G_M$=2.7 is the coupling constant.

The resulting free distribution function, without the phase space corrections
coming from any limited decay parent mass (or total c.m. energy), can be seen
in Fig.~\ref{fig:delta}.

It should be pointed out, that decay of the $\Delta$ resonance in $Ne^+e^-$ is
unmeasured up to now, and the correct treatment is still under
discussion. Moreover, the bremsstrahlung has be taken into account in a
coherent ways, since the final state is equal.

A very promising ansatz has been published recently for the bremsstrahlung and
$\Delta$ case~\cite{kaptari}, and the work to include such processes 
into the Pluto framework
has been started.

\subsection{Vector meson direct decay}
\subsubsection{Calculation of the partial width}

\begin{figure}
\begin{center}
\resizebox{0.59\columnwidth}{!}{%
  \includegraphics{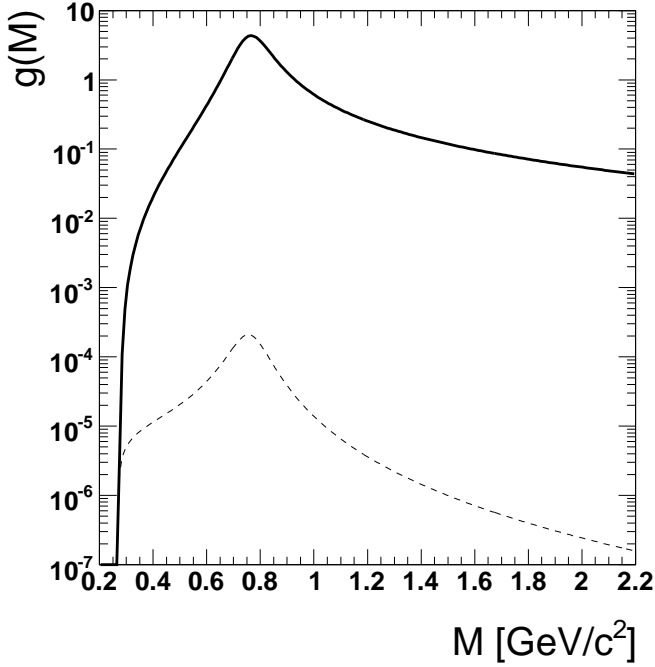}
}
\begin{minipage}[b]{0.40\columnwidth}
\caption{The $\rho^0$ free spectral shape. Solid line: Full shape, dashed line:
  $\rho^0 \to e^+e^-$ with the $\pi\pi$ cutoff as described in the text.
  \newline ~}
\label{fig:rho}       
\end{minipage}
\end{center}
\end{figure}

Vector mesons couple to photons, as it is well known from the Vector Meson
Dominance (VMD) model, and have a direct dilepton decay mode
$\rho^0,\omega,\phi\to e^+e^-$. The decay products (electron-positron) have
obviously fixed masses, therefore mass sampling is not an issue here, but
since this is perhaps the most important process from the point of view of
HADES, the decay widths (and branching ratios) are explicitly calculated in the
code. The mass dependence of the direct vector-meson dilepton decay width for
is given by~\cite{ko}:
\begin{equation}\label{eqn:direct_decay}
  \Gamma^{V\to e^+e^-}(m)
  =\frac{c_V}{m^3}
  \sqrt{1-\frac{4m_e^2}{m^2}}
  \left(
    1+\frac{m_e^2}{m^2}
  \right)
\end{equation}
where the the index $V$ refers to one of $\rho^0,\omega$ and $\phi$, and $c_V$
is $3.079\cdot 10^{-6}$, $0.287\cdot 10^{-6}$, and $1.450\cdot 10^{-6}$~GeV$^4$
respectively~\cite{cite_18_siemens}.

In addition, we follow the ansatz here that the $\rho$ is governed by the
2-Pion phase space in order to be comparable to transport code
calculations~\cite{diana}.  This cut-off behavior at $2\cdot M_{\pi^0}$ can
be seen in Fig.~\ref{fig:rho}. However, this is still a question under
discussion.

\subsubsection{$\rho-\omega$ mixing effect}\label{complex}
The Breit-Wigner Eqn.~\ref{eqn:breit_wigner} can also be replaced by the
absolute value squared
\begin{equation}\label{eqn:complex_bw2}
  g^{\rm k}(m) = \left| \Pi^{\rm k}_0 \right|^2
\end{equation}
of the complex amplitude
\begin{equation}\label{eqn:complex_bw}
  \Pi^{\rm k}_0 = A_0\frac{e^{-{\rm i}\phi} m \sqrt{\Gamma^{\rm k}(m)} }
  {M_{\rm R}^2-m^2+{\rm i}m \Gamma^{\rm tot}(m)}
\end{equation}
which is an alternative approach and can be enabled in Pluto by the user
interface.  This offers the possibility to let the leading term $\Pi^{\rm
  k}_0$ of the decay channel ``k'' to be interfered with a various number of
different terms. The expression for such an interference is the coherent sum
of all contributing terms:
\begin{equation}\label{eqn:complex_bw3}
  \Pi^{\rm k} = \sum_j A_j e^{{\rm i}\phi_j} \Pi^{\rm k}_j 
\end{equation}
among with the relative phase $\phi_j$ and the mixing intensity $A_j$.  The
additional terms $\Pi^{\rm k}_j,\ j=1...n$ can be obtained from single
resonance models, decay models or even an exchange graph which is
not assigned to a fixed decay mode. Pluto is able to add such stand-alone
contributions, which makes it open for more advanced theoretical studies.

As an example for such calculations, the well known $\rho^0-\omega$-mixing has
been implemented, which arises in the coupling of the $\rho^0$-meson to the
$e^+e^-$ channel because the electromagnetic force does not conserve the $I_3$
component of the Isospin. This discussion has been started already some time
ago~\cite{sakurai} and became important in the $ee\to \pi\pi$ scattering
experiments. On the other hand, taking the time-reversal reaction it should play
a role in the di-lepton decay of the $\rho^0$ meson, in particular if produced
in the $\pi\pi$-fusion.

After enabling the complex Breit-Wigner model for both the $\rho^0$ and the
$\omega$, the $\rho^0$ amplitude is calculated to be:
\begin{equation}\label{eqn:complex_bw4}
  \Pi^{\rm \rho^0 \to ee} = \Pi^{\rm \rho^0 \to ee}_0 + A e^{{\rm i}\phi} \Pi^{\rm \omega \to ee}
\end{equation}
with the relative phase $\phi=-1.60$ and the mixing amplitude $A=0.039$.

\begin{figure}
\begin{center}
  \begin{minipage}[b]{0.59\columnwidth}
    \resizebox{\columnwidth}{!}{%
      \includegraphics{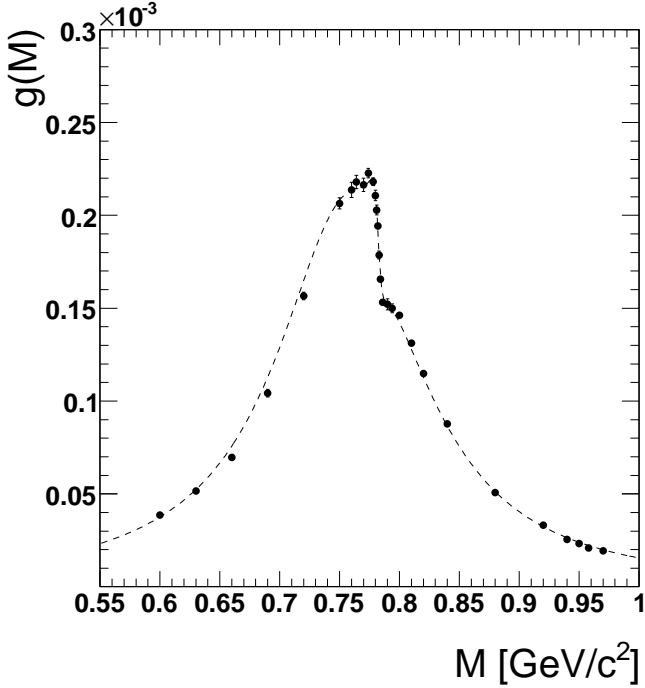}}
  \end{minipage}  
  \begin{minipage}[b]{0.40\columnwidth}
    \caption{Alternative $\rho^0$ spectral function including the $\rho^0 - \omega$ mixing
      (dashed line) compared to the data from~\cite{barkov}. \newline ~}
    \label{fig:rho_sample_compl}       
  \end{minipage}
\end{center}
\end{figure}

It should be noted that this appearing deviation from the normal Breit-Wigner
shape is negligible for the dilepton cocktail and it is used here for test
and demonstration purpose only. 

\section{Polar angle distributions and many-body correlations}\label{angular}

By default, the code samples scattering angles in the rest frame of the parent
particle isotropically. Multi-particle emission (after the mass sampling has
been done as described before) is performed by the Genbod
algorithm~\cite{genbod} which calculates the momenta according to phase space.
For a few select channels, however, empirical parameterizations of angular
distributions and multi-particle correlations have been implemented.

\subsection{$pp$ elastic scattering}

Elastic $pp$ scattering is important for detector and spectrometer calibration
studies. It is therefore useful to have in hand a convenient parameterization
for sampling realistic scattering angles. Pluto includes a parameterization
based on a phase-shift analysis encompassing the world data, from an algorithm
(SAID) supplied by R. Arndt~\cite{said}. This yields elastic pp scattering
distributions accurate to within a fraction of 1\% for proton beam energies
expected for HADES experiments.  It should be noted that the sampled range of
angles in the center of mass is [1,179] degrees, in order to avoid the
singularities at forward and backward angles due to the Coulomb potential.
This is not a limitation since extreme angles practically coincide with the
beam path, where no detection is possible.

\subsection{The $\eta$-case}
\subsubsection{The reaction $pp \to pp \eta$}

\begin{figure}
a.) \resizebox{0.45\columnwidth}{!}{%
  \includegraphics{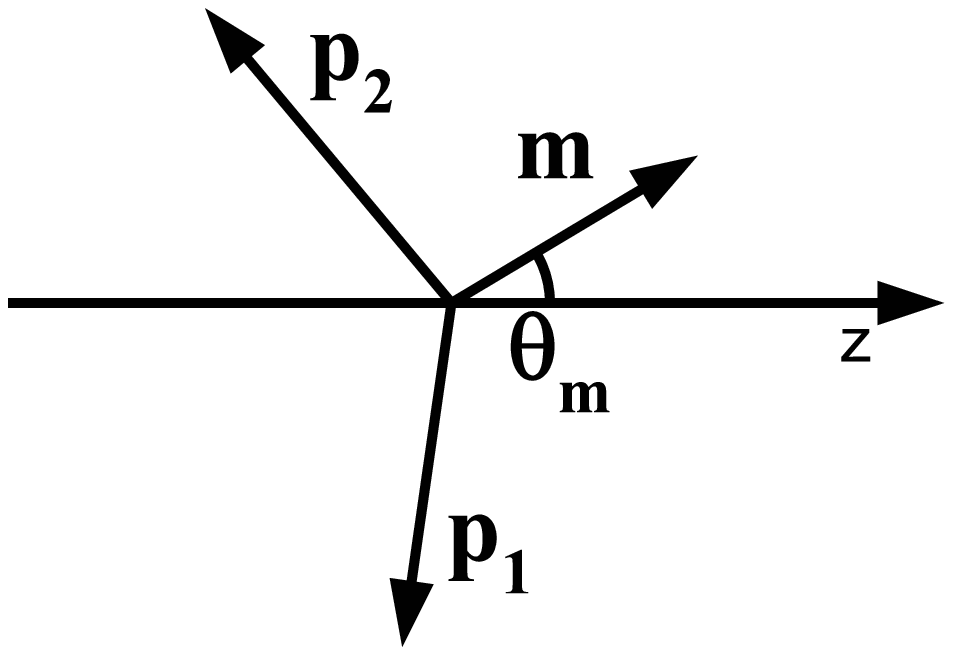}
}
b.) \resizebox{0.45\columnwidth}{!}{%
  \includegraphics{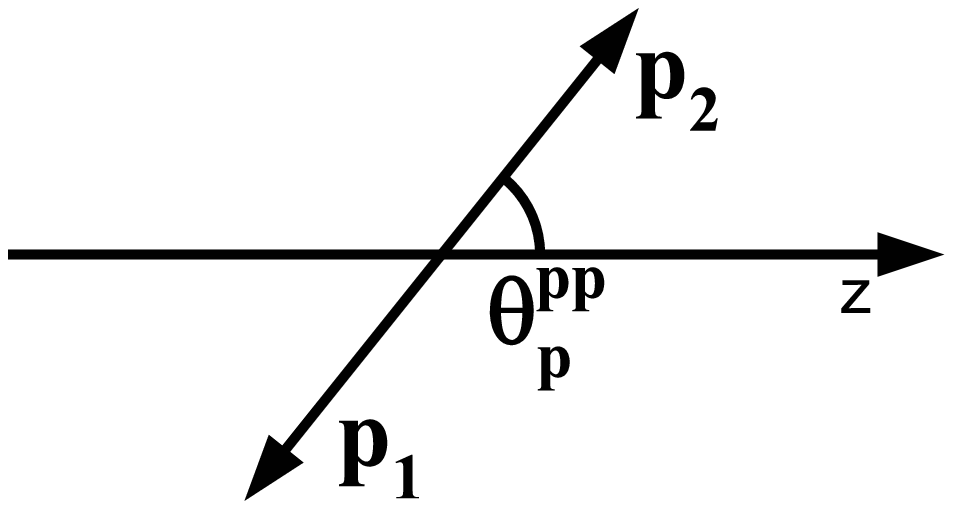}
}
  \caption{Parameterization of the angles in three-body production: a.) 
    The polar angle of the emitted meson $m$, b.) The angular alignment of the
    $pp$ pair.  }\label{fig:meson_polar_angles}
\end{figure}

The DISTO collaboration has reported an anisotropy of the $\eta$ polar angle
$\theta^{c.m.}_\eta$ in the $pp$ collision~\cite{disto_eta} at 2.15, 2.5 and 2.85~GeV
beam energy.  Closer to threshold, this anisotropy seems to
vanish~\cite{cosy-tof}.  In addition, the proton angular alignment (which is
the distribution the polar angle $\theta^{pp}_p$ of any proton in the common $pp$ rest
frame, with respect to the beam momentum) tends to be aligned stronger in
forward/backward direction with increasing beam energy.
Fig.~\ref{fig:meson_polar_angles} sketches these angles.

Such behavior is usually described with a fit using Legendre-polynomials. In
the $\eta$ case, the usage of the first 2 even Legendre-polynomials have found
to be sufficient and hence the differential cross section can be parameterized as:

\begin{equation}
\frac{d\sigma}{d\Omega} \propto 1 + c_2 \cdot \frac{1}{2}\left(3  \cos^2 \theta -1 \right)
\end{equation}
where a fit to the given data points has been applied using a 2nd order
polynomial as a function of the total c.m. energy $Q$ in the $pp$ system:

\begin{equation}\label{c2_param}
c_2 = a_0 + a_1 Q + a_2 Q^2
\end{equation}
Obtained by this method, the following values have been used:

\begin{description}

\item[$\eta$-case:] $a_0=37.4$, $a_1=-27.7$, $a_2=5.07$
\item[$pp$-case:] $a_0=5.04$, $a_1=-4.54$, $a_2=1.01$

\end{description}

\subsubsection{Matrix element in $\eta\to\pi^+\pi^-\pi^0$}

The Dalitz plane of the $\eta$ decay into 3 charged pions shows a strong
non-phase space behavior, which is caused by the difference of the light
quark masses~\cite{leutwyler}. The decay slopes of this plane are usually
parameterized as:
\begin{equation}
  x=\sqrt{3}\frac{T_{\pi^-}-T_{\pi^+}}{Q}, y=3\frac{T_{\pi^0}}{Q}-1
\end{equation}
with $T_\pi$ the kinetic energy of the individual pion in the $\eta$ rest
frame and $Q=m_\eta-2m_{\pi^+}-m_{\pi^0}$. Then, the matrix element (which is
the deviation from the constant value of the Dalitz plot) can be parameterized
as $1+ax+by+cxy$. Pluto includes
the result from Crystal
Barrel~\cite{cbarrel1} which is $a=-9.94, b=0.11, c=0$.

\subsection{Production of $\omega$ mesons in Pion-induced reactions}

This reaction $\pi N \to N \omega$ follows Ref.~\cite{cite_13_omega}, where the
angular distribution for $\omega$ production in $\pi N \to N \omega$ in the
center-of-mass frame was found to be sharply (exponentially) peaked at forward
angles, consistent with the parameterization
\begin{equation}\label{eqn:omega_production}
  f(\cos\theta_\omega^{c.m.})=1+\alpha e^{-4(1-\cos\theta)}
\end{equation}
where $\alpha$ depends on the invariant mass, and has been parameterized as
\begin{equation}\label{eqn:omega_production2}
  \alpha(\sqrt{(s)})=6.9\times10^{-3}e^{2.873\sqrt{s}}
\end{equation}
by fitting the data of Ref.~\cite{cite_13_omega}. In a similar ways, the 
$\pi^+ + p \to \Delta^{++} + \omega$ and $\pi^+ + p \to \pi^+ + p + \omega$ reactions
have been included, using the data from~\cite{alff}.

\subsection{The $\Delta (1232)$ case}

\subsubsection{Angular distribution in the $NN \to N\Delta (1232)$ production}

This follows Ref.~\cite{cite_13_dmitriev}, which is in excellent agreement
with the data. The direct and exchange matrix elements,  averaged over all the
spin states, are given by
\begin{equation}\label{eqn:delta_decay_angles}
  \begin{array}{lll}
    \frac{1}{4}\sum_{\lambda_1\lambda_2\lambda_3\lambda_4}
    \left|M({\rm direct})\right|^2
    & = &
    \left( \frac{g_\pi f_\pi^*}{m_\pi}\right)
    \frac{F^4(t,m)}{t-m^2_\pi}
    t \left[t-(m-m_N)^2\right]
    \frac{\left[t-(m+m_N)^2\right]^2}{3m^2}
  \end{array}
\end{equation}
\begin{equation}\label{eqn:delta_decay_angles2}
  \begin{array}{lll}
    \frac{1}{4}\sum_{\lambda_1\lambda_2\lambda_3\lambda_4}
    \left(M_a^+M_b + M_b^+M_a \right)
    & = &
    \left( \frac{g_\pi f_\pi^*}{m_\pi}\right)
    \frac{F^2(t,m)F^2(u,m)}{(t-m_\pi ^2)(u-m_\pi ^2) }
    \frac{1}{6m^2} \cdot \\ ~\\& & 
    \left[tu+(m^2-m^2_N)(t+u)-m^4+m^4_N\right]  \times \\ ~\\& & 
    \left[tu+m_N(m+m_N)(m^2-m^2_N)\right]\cdot \\ ~\\& &
    \left[tu-(m^2+m^2_N)(t+u)+(m+m_N)^4\right]\cdot \\ ~\\& &
    \left[tu-m_N(m-m_N)(m^2-m^2_N)\right]
  \end{array}
\end{equation}
where $f_\pi=1.008$ and $f_\pi^*=2.202$ are the $\pi N$ and $\pi\Delta(1232)$
coupling constants with all the particles assumed on-shell, $u$ and $t$ are the
standard Mandelstam variables, $m$ is the resonance mass as sampled from
Eqn.~(\ref{eqn:breit_wigner}), and $F(t,m)$ is the mass-dependent form factor
\begin{equation}\label{eqn:delta_decay_angles3}
  F(t,m)=F(t)\sqrt{\frac{M_\Delta}{m}}\frac{\beta^2+(q^\Delta_{m_\Delta m_\pi m_p})^2}{\beta^2+(q^\Delta_{m_\pi m_p}(m))^2}
\end{equation}
where $\beta= 300$~MeV, $q^\Delta_{m_\Delta m_\pi m_p}$ and $q^\Delta_{m_\pi
  m_p}(m)$ are as in Eqn~(\ref{eqn:width_monitz}) the momentum if the frame of
  the resonance (the first one at pole mass), and
\begin{equation}\label{eqn:delta_decay_angles4}
  F(t)=\frac{\Lambda^2-m^2}{\Lambda^2-t}
\end{equation}
is the unmodified form-factor, with $\Lambda=0.63$~GeV from fits to
data~\cite{cite_13_dmitriev}. The argument $t$ in the form factor is one of the
Mandelstam variables. Likewise, exchanging $t$ with $u$ one gets $F(u,m)$ in the
matrix elements above. The mass dependence in the form factor is slightly
modified from the corresponding expression of Ref.~\cite{cite_13_dmitriev},
the difference coming
from the mass-dependent width of Eqn.~(\ref{eqn:width_monitz}).  These matrix
elements result in the following expression for the differential cross section
\begin{equation}\label{eqn:delta_decay_angles5}
  \frac{d\sigma}{dt}=\frac{1}{64\pi} |M|^2\frac{1}{4I^2},\ I=\sqrt{(p_1p_2)^2-M_N^4}
\end{equation}
$I$ is a kinematical factor with $p_{1,2}$ the beam- and target-proton 4-vectors,
and $M_N$ the nucleon mass. 
In switching from $d\sigma/dt$ of Eqn.~(\ref{eqn:delta_decay_angles5}) to
$d\sigma/d\Omega$, actually sampled by the code when scattering angles are
picked, the extra phase space factor $q^{H_p}(m,m_1,m_2)$ introduced in
Eqns.~(\ref{eqn:m1_width}-\ref{eqn:m2_width}) arises naturally. Thus, sampling
the mass $m$ and subsequently the center-of-mass scattering angle from
Eqn.~(\ref{eqn:delta_decay_angles5}) yields spectra consistent with
differential cross sections, in good agreement with experimental data. The
scattering-angle $\theta^{\rm c.m.}_\Delta$ dependence enters through the
Mandelstam variables in the matrix elements.

\begin{figure}
\begin{center}
  \begin{minipage}[b]{0.49\columnwidth}
    \resizebox{\columnwidth}{!}{%
      \includegraphics{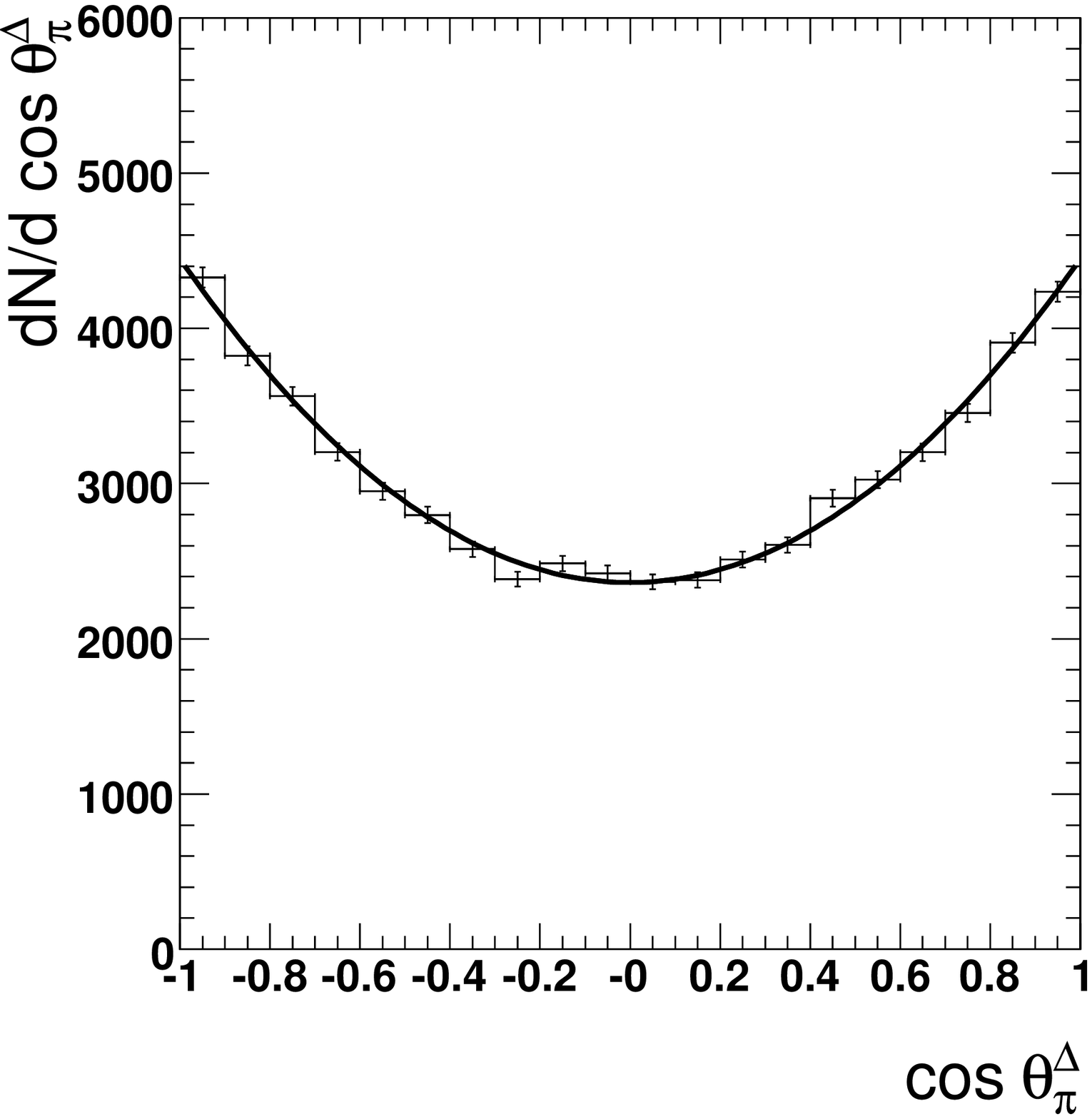}}
  \end{minipage}  
  \begin{minipage}[b]{0.49\columnwidth}
    \resizebox{\columnwidth}{!}{%
      \includegraphics{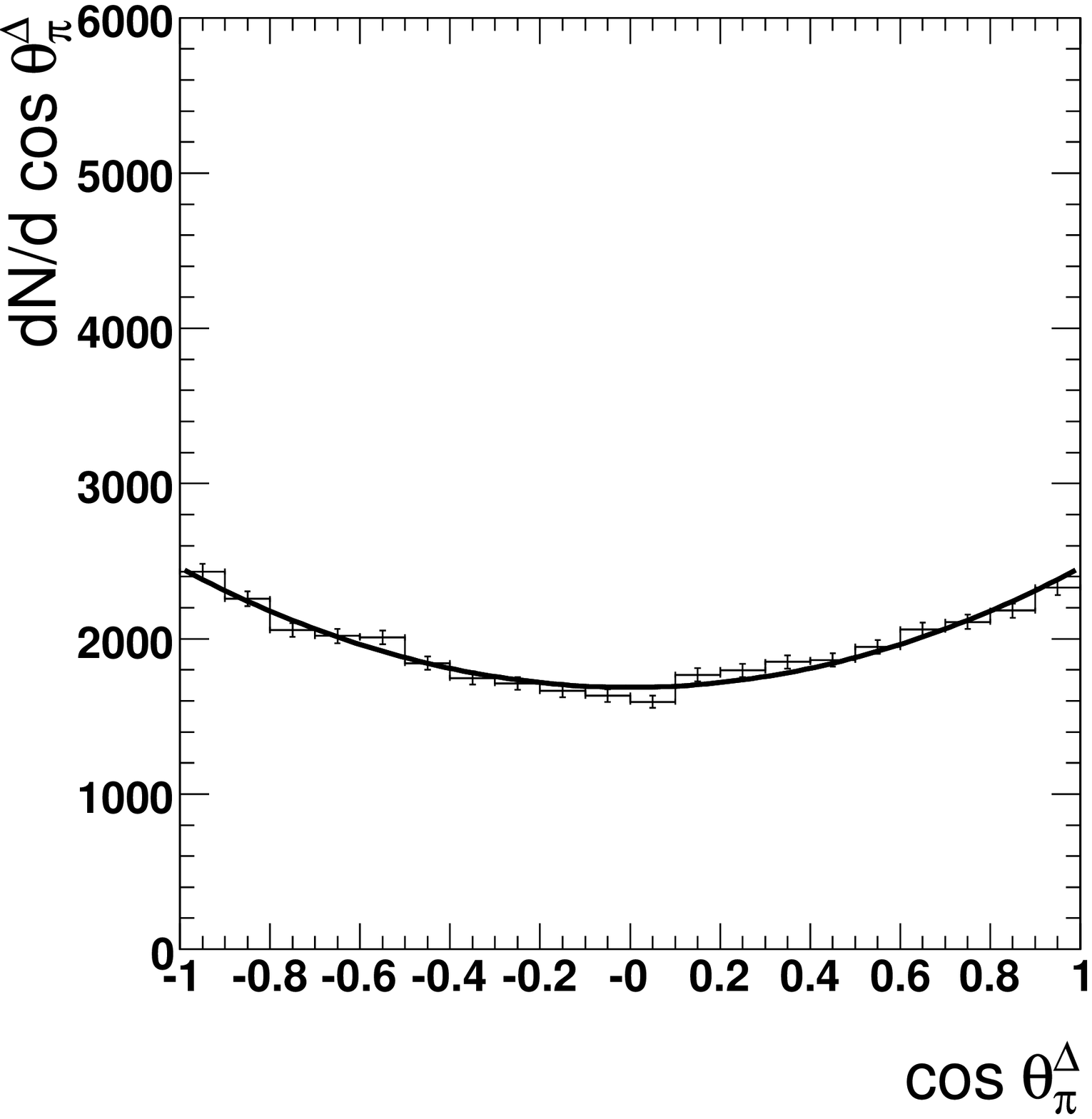}}
  \end{minipage}  
  \caption{ $\cos^2\theta^\Delta_\pi$ distribution of the $\Delta$ decay angle $\theta$ in its
    rest frame and with respect to its c.m. momentum. Left: selection on the
    $\Delta$ emission angle $|\cos^2 \theta^{\rm c.m.}_\Delta|>0.9$, right:
    $|\cos^2 \theta^{\rm c.m.}_\Delta|<0.9$. It is clearly visible that $\Delta$
    production is dominated by forward angles. The fit on the angular
    distribution gives $A=0.88 \pm 0.003$ (left) and $A=0.45 \pm 0.02$
    (right).  }
  \label{fig:delta_angles}       
\end{center}
\end{figure}

\subsubsection{$\Delta (1232) \to N \pi$}\label{pw_delta}

Historically, the partial wave contributions of resonances have been found by
scattering Pions off nucleons and the measurement of the outgoing Pion angular
distribution in the resonance rest frame. In $pp$ collisions, however, the
initial situation is different. The usual picture of such a reaction is the
exchange of mesons (one boson exchange, OBE).

The $\Delta(1232)$ production in $pp$ reactions is well described by pion
exchange amplitudes. However, the decay angle in the $\Delta(1232)$ rest
frame, expect to be $1+A\cos^2\theta^\Delta_\pi$, with $A=3$, is a more
complicated issue. On one hand, the direction of the virtual pion depends on
the question if the direct or exchange term is dominating.  Measurements on
the Treiman-Yang angle suggest the exchange of virtual Pions~\cite{bacon}, but
on the other hand, the extracted coefficient $\bar{A}$ (averaged over all
terms) gave $\bar{A} \approx 0.85$ for the region $\cos^2(\theta^{\rm
  c.m.}_\Delta)>0.9$ and drops down to $\bar{A} \approx 0.41$ for the region
$\cos^2(\theta^{\rm c.m.}_\Delta)<0.8$. Ref.~\cite{wicklund} extracts the
coefficient $\bar{A}$ to be 0.65.  Our approach is to use the polar axis taken
as the momentum transfer direction to the excited nucleon in the c.m. frame by
using the calculation from the previous section, and scale $A$ such that the
result of~\cite{wicklund} is reproduced. This is also consistent with the data
from~\cite{bacon} (see Fig.~\ref{fig:delta_angles}).

\subsubsection{$\Delta (1232) \to N \gamma$}

For the production of virtual photons, one has to calculate the
photoproduction amplitudes (for a very good introduction on this topic
see~\cite{Napolitano}). Assuming only the $M_{1+}$ transition (so the spin
flip of one single quark), which is a good approximation in the
$\Delta(1232)$-case, the expected distribution of the Pion in the
photoproduction $\gamma N \to N \pi$ is to be $5-3\cos^2\theta_\pi^\Delta$.
With time-reversal arguments, one expects that the virtual photon angular
distribution shows the same behavior which is our approach taken but the same
damping factor has been used as obtained in the $p\pi$ case.

\subsection{Quasi-free scattering}

The quasi-free scattering of a nucleon $N_1$ on a nucleus $A$ (or vice versa)
is considered in Pluto in two steps: First, the Fermi-momentum and off-shell
mass of the nucleon $N_2$ inside the nucleus $A$ are determined and the
particle properties are set up correctly, and in a second step the reaction
$N_1 + N_2$ is performed with all the consecutive decay modes as defined by
the user.  At the moment a dedicated sampling model is included for the
deuteron wave function~\cite{benz}.

\section{Thermal sources}\label{thermal}

Thermal sources are needed in the case of heavy ion reactions~\cite{hades_prl} in
order to extrapolate the meson production yield measured with HADES to the full solid angle
and to subtract the cocktail of ``trivial'' sources from the measured
di-lepton spectrum.

In this case, Pluto is able to emit mesons and baryonic resonances without a
collision and without considering energy and momentum conservation, by a
special particle which we call ``fireball''. This means that first the
particles are created (for each particle species we set up one fireball) and
sub-sequentially decay in the Pluto framework as described before. This
systematic procedure allows directly comparing the elementary reactions with
results obtained in heavy ion reactions.

In the case of stable (long-lived) particles only the total energy $E$ is
sampled as a relativistic Boltzmann distribution in the nucleus-nucleus c.m.
frame:
\begin{equation}\label{boltzmann}
  \frac{dN}{dE}\propto p\ E \ e^{-E/T}
\end{equation}
This distribution is not explicitly normalized to 1, this is done numerically
by the ROOT TF1 object.
A source with two temperatures, as observed e.g. in Pion production
is realized by:
\begin{equation}\label{boltzmann2}
  \frac{dN}{dE}\propto p\ E \ \left[ fe^{-E/T_1} + (1-f)e^{-E/T_2} \right]
\end{equation}
where $f$ and $1-f$ are the respective fractions of the two components.
Optionally, radial flow is implemented using the Siemens-Rasmussen
formulation~\cite{cite_18_siemens}\footnote{Note the typos in Eqn. 1,
  corrected e.g. in~\cite{reisdorf}.}:
\begin{equation}\label{radial_flow}
  \frac{dN}{dE}\propto p\ E\ e^{-\gamma_r\frac{E}{T}}
  \left[\left(\gamma_r+\frac{T}{E}\right)\frac{\sinh
      \alpha}{\alpha}-\frac{T}{E}\cosh \alpha \right]
\end{equation}
with
\begin{itemize}
  \item $\beta_r$: Blast velocity
  \item $\gamma_r = 1/\sqrt{1-\beta_r^2}$
  \item $\alpha=\beta_r \gamma_r p/T$
  \item $p=\sqrt{E^2-m^2}$
\end{itemize}
Note that in the limit $\beta_r \to 0$, Eqn.~(\ref{boltzmann}) is
recovered. In case of two temperatures ($T_1,\ T_2$), Eqn.~(\ref{radial_flow})
is extended as:
\begin{equation}\label{radial_flow2}
  \begin{array}{cc}
    \frac{dN}{dE}\propto & p\ E\ \left\{f 
        e^{-\gamma_r\frac{E}{T_1}}
        \left[\left(\gamma_r+\frac{T_1}{E}\right)\frac{\sinh
            \alpha_1}{\alpha_1}-\frac{T_1}{E}\cosh \alpha_1 \right]\right. \\ &
        +\left.(1+f)e^{-\gamma_r\frac{E}{T_2}}
        \left[\left(\gamma_r+\frac{T_2}{E}\right)\frac{\sinh
            \alpha_2}{\alpha_2}-\frac{T_2}{E}\cosh \alpha_2 \right]
      \right\}
  \end{array}
\end{equation}
These distributions are sampled spatially isotropic or, optionally, with:
\begin{equation}\label{ang_hi}
  \frac{dN}{d\Omega} \propto 1+ A_2 \cos^2 \theta^{\rm c.m.} + A_4 \cos^4\theta^{\rm c.m.} 
\end{equation}
Note that most transport models and some data too show an
dependence of the temperature on the angle, thus $T=T(\theta^{\rm c.m.})$. Such an
effect can be optionally and roughly modeled in Pluto as well.

\begin{figure}
\begin{center}
  \begin{minipage}[b]{0.49\columnwidth}
    \resizebox{\columnwidth}{!}{%
      \includegraphics{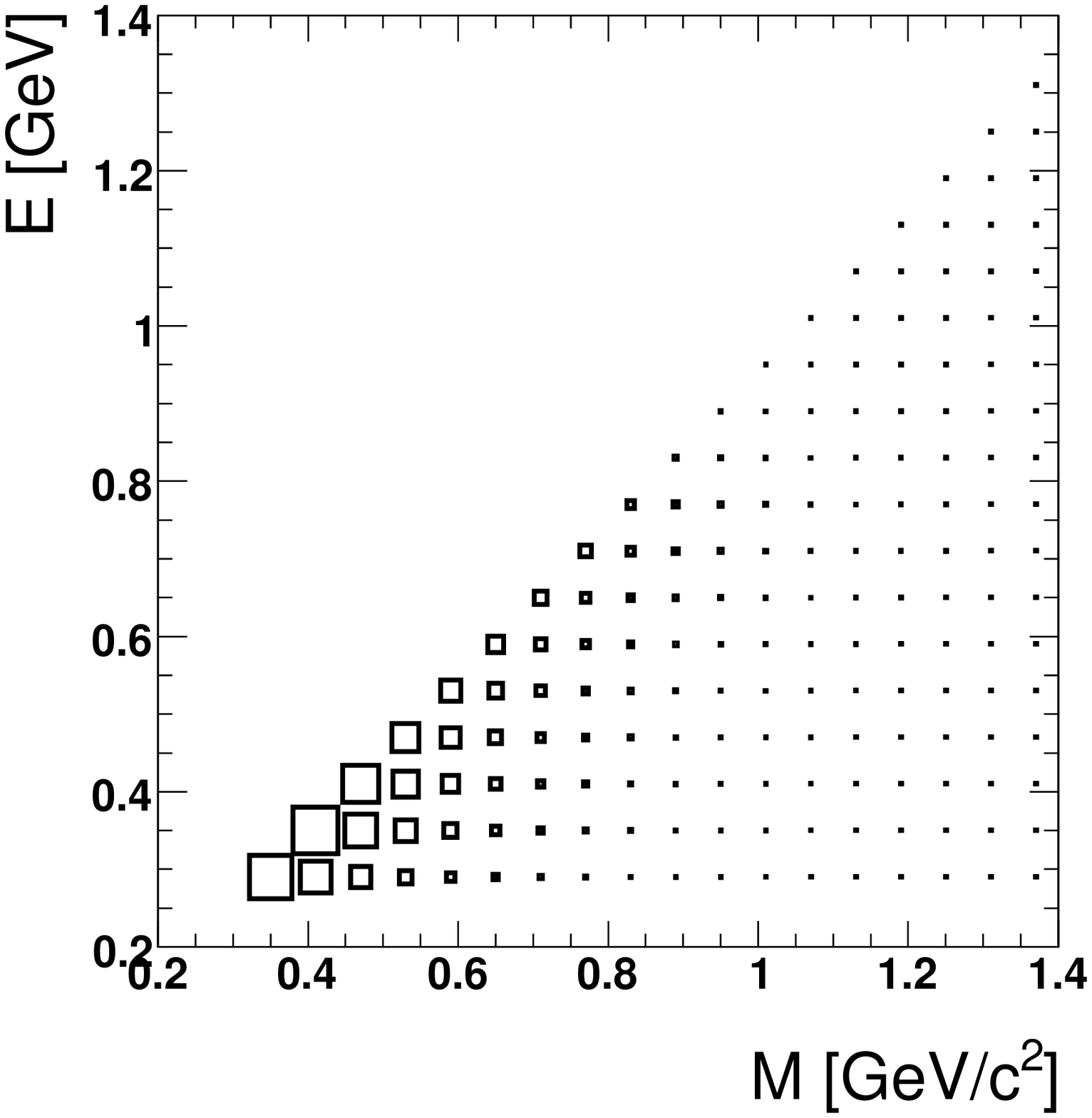}}
  \end{minipage}  
  \begin{minipage}[b]{0.49\columnwidth}
    \resizebox{\columnwidth}{!}{%
      \includegraphics{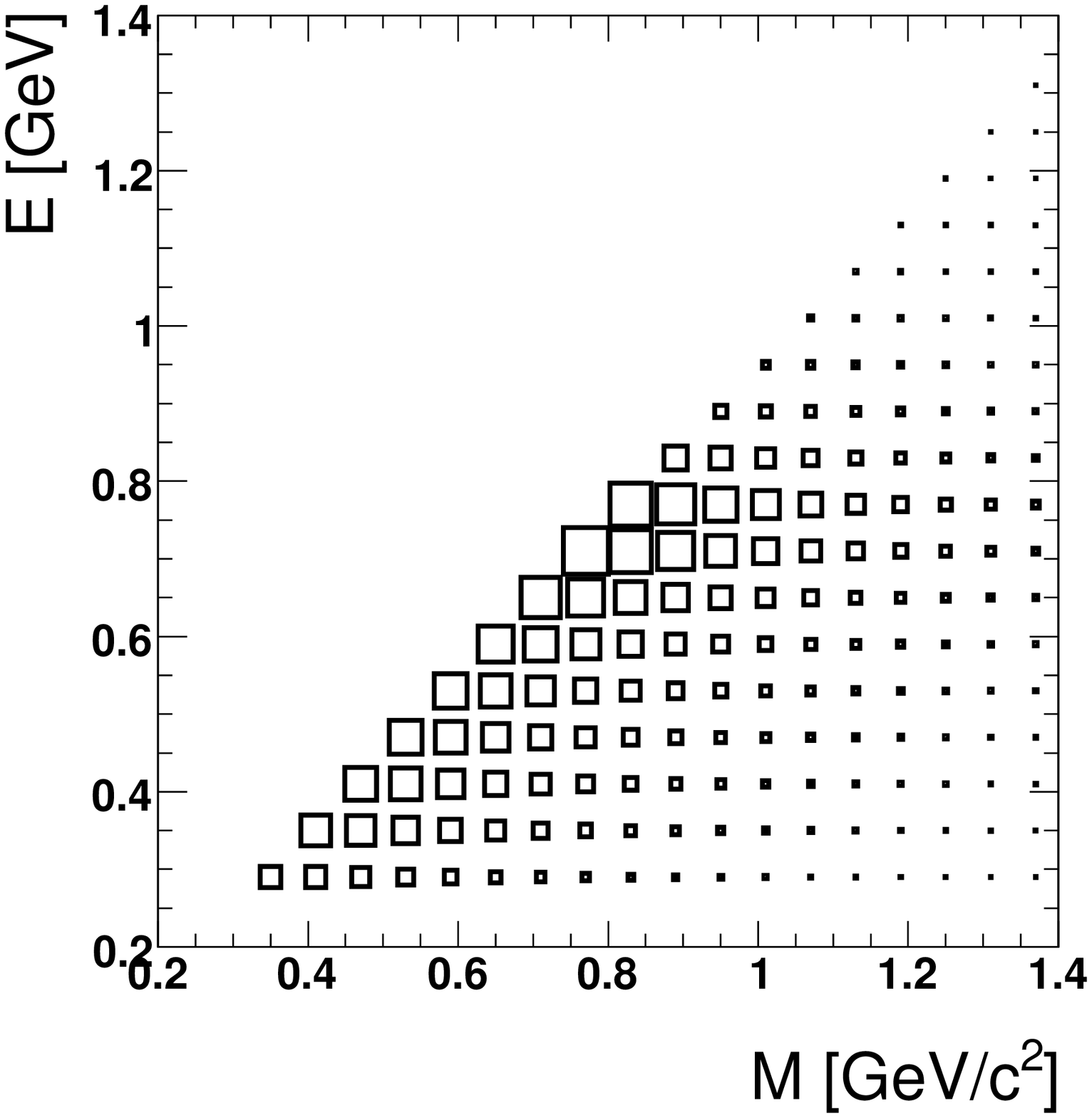}}
  \end{minipage}  
  \caption{Pluto generated $d^2/dNdM$ distributions for the leptonic $\rho^0$
    at $T=0.1$~GeV (left) and $T=0.05$~GeV (right).
  }
  \label{fig:thermal_rho}       
\end{center}
\end{figure}

\begin{figure}
\begin{center}
  \begin{minipage}[b]{0.59\columnwidth}
    \resizebox{\columnwidth}{!}{%
      \includegraphics{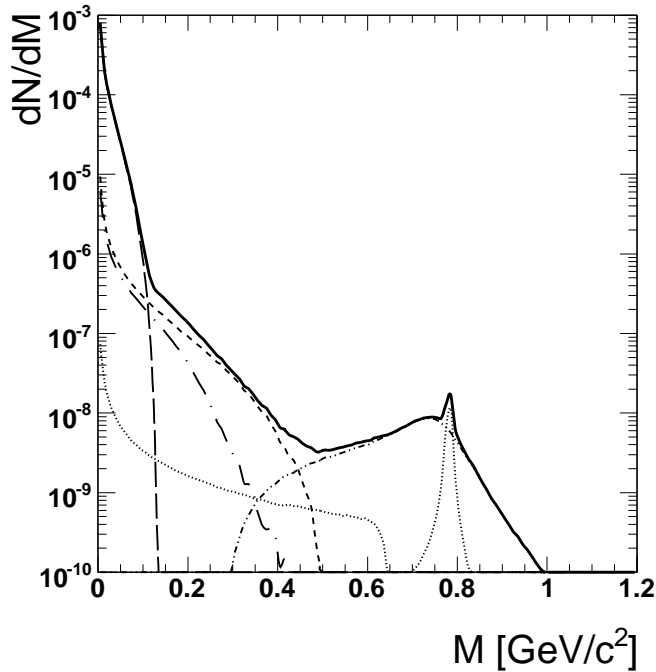}}
  \end{minipage}  
  \begin{minipage}[b]{0.40\columnwidth}
    \caption{Thermal dilepton spectrum in the $^{12}$C+$^{12}$C reaction at 2~GeV per nucleon.
      Solid line: all contributions, long dashed line: $\pi^0$ Dalitz decay, 
      short dashed line: $\eta$ Dalitz decay,
      dashed-dotted line: $\Delta$ Dalitz decay,
      dotted line: $\omega$ decay (direct and Dalitz),
      dashed-dotted-dotted: $\rho^0$ direct decay.
      \newline ~}
    \label{fig:c2c_inv}       
  \end{minipage}
\end{center}
\end{figure}

For broad particles, the energy and mass are sampled as:
\begin{equation}\label{boltzmann3}
  \frac{d^2N}{dEdm} \propto \frac{dN}{dE}\cdot  g(m) \cdot \theta(E > m)
\end{equation}
with $g(m)$ from Eqn.~(\ref{eqn:breit_wigner}). In the case that only one
selected subsequent decay k should be calculated, $g(m)$ has to be replaced
with $g^{\rm k}(m)$ which includes the mass-dependent branching ratios.

This distribution depends strongly on the temperature. Examples are shown in
Fig.~\ref{fig:thermal_rho} for $T=0.1$~GeV and $T=0.05$~GeV.

Fig.~\ref{fig:c2c_inv} shows a final dilepton final spectrum without any
detector acceptance effects, where the parameters of the fireball and the
relative abundances of the sources are adjusted to the reaction
$^{12}$C+$^{12}$C~\cite{hades_prl}.

\section{User interface for event production}\label{interface}

After the features of the Pluto package have been presented, the user interface
and technical implementation of the framework are roughly described. The
package does not need any additional libraries (beside ROOT~\cite{root}),
which makes it possible to use it as a standalone environment for quick
detector studies. Moreover, it is small and fast (a typical reaction with 1
million events takes only some few minutes). On the other hand, it is very
important for advanced studies to allow the user for changes of almost all
parameters and including new ideas. The latter one can be done even without
recompilation, which is supported by the smooth interaction of the
ROOT-interpreter with C++. An interface to attach different input sources and
to allow 3$^{\rm rd}$ party event generators for interaction is available as
well.

\subsection{Basic components}

The main objects of Pluto are \classref{PParticles}. This class defines
``particle'' objects, the most elementary unit in the context of simulations
with this package, and contains functions for handling particle observables.
In the parlance of C++, this class inherits from native 3- and 4-vector ROOT
classes~\cite{root}.

\begin{figure}
\begin{center}
\resizebox{0.6\columnwidth}{!}{%
  \includegraphics{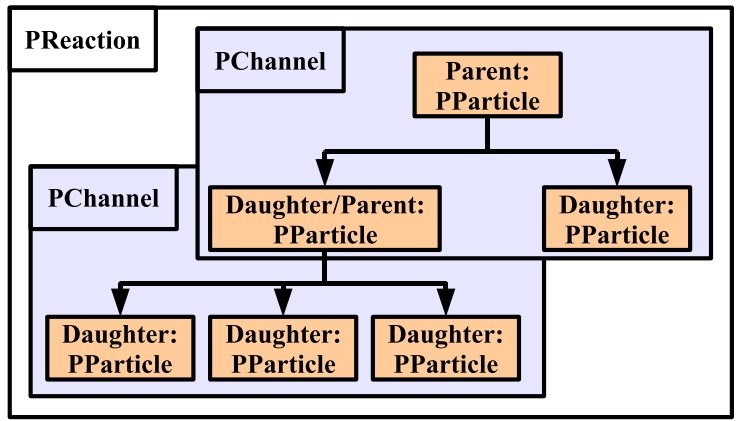}
}
  \caption{Basic classes of the Pluto interface: A single decay step (of one
    parent into several daughters) are combined in a \classref{PChannel}. The
    complete \classref{PReaction} is formed by consecutive
    \classref{PChannels}.}\label{fig:pchannel}
\end{center}
\end{figure}

These objects are modified via \classref{PChannels}, which handle the decay of
one ``parent'' particle into several ``daughter'' particles, respectively.
Fig.~\ref{fig:pchannel} is sketching this basic concept.  During runtime,
these \classref{PChannels} are connected to a \classref{PDistributionManager}
which offers a list of included distributions and models for coupled-channel
calculations and thus much of hadronic interaction models, like the empirical
angular-distribution parameterizations the mass and momentum sampling.

Since particle production is done via a complete chain of consecutive decays,
several \classref{PChannels} are finally connected in a \classref{PReaction},
which also contains functions for the execution of simulated event loops (see
also Fig.~\ref{fig:pchannel}).

Last, multi-reaction cocktail calculations are facilitated via the
\classref{PDecayManager}.

The following sub-sections describe how to set up a reaction, starting from a
very simple example for the first usage, up to the description of the advanced
features of Pluto. All examples are based on ROOT-macros (or interactive
sessions), which is a very elegant way to control and steer the Pluto package.

\subsection{Getting started}

The simplest approach to perform a reaction is to open a ROOT-CINT session
and use the reaction parser of Pluto:

{\footnotesize
\begin{verbatim}
> PReaction my_reaction(6,"p","p","p p rho0 [e+ e-]","rho_sample",1,0,0,0);
> my_reaction.loop(100000);
\end{verbatim}}
  
  The arguments of the constructor which is used here, are the beam momentum,
  the beam and target particle name, and the decay products, separated by
  spaces (including their decay in brackets), the filename, and 4 flags which
  are described in detail in Sec.~\ref{preaction}.  The second command in this
  example actually opens the ROOT file "rho\_sample" and produces 100000
  events. In each of these events, $\rho^0$ mesons are sampled in $pp$
  collisions at 6.0~GeV beam momentum using three body phase space scaling and
  a relativistic Breit-Wigner for the broad $\rho^0$ meson, as already
  discussed in Sec.~\ref{m3_model} (e.g. Fig.~\ref{fig:rho_sample_phase}).
  This included physics can be shown in each case with the
  \classref{Print()}-method of the \classref{PReaction} class:

{\footnotesize
\begin{verbatim}
> my_reaction.Print();
  Reaction of 6 Particles interacting via 2 Channels
   Reaction Particles:
     0. quasi-particle (fixed p beam and p target)
     1. p
     2. p
     3. rho0
     4. e+
     5. e-
   Reaction Channels:
     1. p + p --> p + p + rho0
        Interaction model(s):
        [p + p_m3_p_p_rho0] 3-body phase space rho0 <PBreitWigner>
        [p + p_genbod_p_p_rho0] Pluto build-in genbod
     2. rho0 --> e+ + e-
        Interaction model(s):
        [rho0_ee_e-_e+] Dilepton direct decay
        [rho0_genbod_e-_e+] Pluto build-in genbod
   Output Files:
     Root : rho_sample.root, all particles on file.
\end{verbatim}}

  This exhibits that the reaction involves 6 particles and is performed in 2
  steps which are the production and the consecutive decay of the
  $\rho^0$-meson. The identifiers in the brackets can be used for the unique
  identification of the interaction models (for details
  see Sec.~\ref{pdistributionmanager}).


  
  The resulting $\rho^0$-meson (with particle id 42) mass can be analyzed by
  re-opening the ROOT file and projecting the mass to a histogram: {\footnotesize
\begin{verbatim}
data.Draw("M()","ID()==41");
\end{verbatim}}

Decays can also be nested, like in the case $\eta\to\gamma^*\gamma\to e^+e^-\gamma$:

{\footnotesize
\begin{verbatim}
PReaction my_reaction(3.13,"p","p","p p eta [dilepton [e+ e-] g]",
    "eta_dalitz",1,0,0,0);
my_reaction.Print();   //The "Print()" statement is optional
\end{verbatim}}
Virtual photons are named ``dilepton'' or ``dimuon'' in Pluto, depending on
the final decay.

\subsection{Complete example}

A single reaction may also be defined by instantiating the objects by hand, which is
demonstrated using the same reaction as above:

{\footnotesize
\begin{verbatim}
PParticle p1("p",0.,0.,3.13);
PParticle p2("p");
  
PParticle q=p1+p2;       //construct the beam particle
  
// eta production
PParticle p3("p");
PParticle p4("p");
PParticle eta("eta");
PParticle *eta_part[]={&q,&eta,&p4,&p3};
PChannel eta_prod(eta_part,3,1);
  
// eta dalitz decay
PParticle di_eta("dilepton");
PParticle g_eta("g");

PParticle *dalitz_part_eta[]={&eta,&di_eta,&g_eta};
PChannel dalitz_decay_eta(dalitz_part_eta,2,1);
  
// decay of the eta dilepton   
PParticle em_eta("e-");
PParticle ep_eta("e+");
PParticle *dileptons_eta[]={&di_eta,&em_eta,&ep_eta};
PChannel dilepton_decay_eta(dileptons_eta,2,1);
  
PChannel *c[ ]={&eta_prod,&dalitz_decay_eta,&dilepton_decay_eta};
PReaction r(c,"eta_dalitz",3,0,0,0,1);
r.loop(10000);
\end{verbatim}}

The structure of this macro becomes more clear after the next sub-sections,
where the classes are described in more detail.

\subsection{PParticle}\label{pparticle}
A \classref{PParticle} is a Lorentz vector, together with a particle id (pid)
and a weight.  The pid convention in the \classref{PParticle} class is
consistent with GEANT3~\cite{geant}, except for the additional unstable
particles of Pluto. The weight is unity by default, unless explicitly set
otherwise, and is updated self-consistently depending on the physics model of
the interaction that produces a given particle. Composite particles made up of
two (but not more) constituent particles may be defined, where the pid
assignment follows the ansatz pid = pid1*1000 + pid2 for the two constituent
pid's. The ``addition'' is used for this operation, intended for the creation
of a quasi particle at the entrance channel from the interaction of a beam
particle (1st constituent) with a target (2nd constituent). For composite
particles, the 4-vector is the sum of the constituent 4-vectors, and the weight
is the product of the constituent weights (uncorrelated weights assumed).

Several constructors are available for instantiating particles, where the last
2 have been used in the example above:

\begin{itemize}
\item \classref{PParticle(char * id, double T);}
\item \classref{PParticle(char * id);}
\item \classref{PParticle(char * id, double px, double py, double pz);}
\end{itemize}

Here, ``id'' is the unique particle name and $T$ the particle kinetic
energy. $px,px,pz$ are the 3-momentum components of the particle.
Functions to return physical observables such as the momentum, velocity- or
Lorentz- vector, weight, rapidity, mass, angles are inherited from the parent
\classref{TLorentzVector} class.

Particles participating in a simulation are instantiated in advance and
subsequently updated during the execution of an event loop. In this way,
unnecessary invocation of time-consuming constructors and destructors is
avoided. The masses are reassigned automatically via
sampling if appropriate, e.g. for unstable resonances, during the execution of
an event loop.

An example of declaring a proton is shown, passing the kinetic energy (GeV) as
argument (with the momentum assumed along the z axis): 

{\footnotesize
\begin{verbatim}
> PParticle p("p",2.);    // proton with kinetic energy (GeV)
> p.Print();              // print member function
  p (0.000000,0.000000,2.784437;2.938272) wt = 1.000000, m = 0.938272 pid = 14
  Vertex = 0.000000 0.000000 0.000000
\end{verbatim}}
  
  An example of a composite particle is illustrated next:

{\footnotesize
\begin{verbatim}
> PParticle p1("p",0,0,2);   // p with 2 GeV/c momentum along the z-axis
> PParticle p2("p");         // proton at rest (default constructor)
> PParticle q=p1+p2;         // composite particle: beam + target (in this
>                            // order)
  Info in <PParticle::operator+>: (ALLOCATION) Keeping beam and target 
      particle for further reference
  Info in <PParticle::operator+>: (ALLOCATION) The composite p + p has been 
      added
> q.Print();                 // displays info
  quasi-particle (0.000000,0.000000,2.000000;3.147425) wt = 1.000000, 
      pid1 = 14, pid2 = 14
  Vertex = 0.000000 0.000000 0.000000
> cout << q.ID() << endl;    // composite particle pid convention: pid2*1000 
                             // + pid1
  14014                      // quasi-particle pid
\end{verbatim}}
  
  This demonstrates, that Pluto adds new composite particles in the data base
  by a background operation whenever occurring, and the original scattering
  particles are kept for later calculations. More composite particles are
  created likewise in case a quasi-free reaction is studied, e.g. the
  scattering of $p+n$ in the $p+d$ reaction. Here, first the momentum of the
  quasi-free nucleon is sampled and consequently the composite particle is
  updated in the event loop.

The particle properties and decay channels can be obtained via the following
command:
{\footnotesize
\begin{verbatim}
> makeStaticData()->PrintParticle("dilepton")    
  Primary key=52
  Primary name=dilepton 
  Pluto particle ID=51 
  Particle static width [GeV]=0.000000 
  Particle pole mass [GeV]=0.001022 

  This particle decays via the following modes:
  Primary key=160
  Primary name=dilepton --> e+ + e- 
  Decay index=90 
  Branching ratio=1.000000 
  Decay product 1->Primary name=e- 
  Decay product 2->Primary name=e+ 
\end{verbatim}}
  
  This gives a simple method to read the information from the Pluto data base,
  where the properties of all particles and decays are stored. More examples
  using the data base, including the addition of new particles and decays can
  be found in Sec.~\ref{database}.

\subsection{PChannel}\label{pchannel}

A \classref{PChannel} is a single step in a reaction process, consisting of a parent,
elementary or quasi-particle from a beam-target interaction, and its
subsequent decay into a number of decay products via a specified decay
mode. The \classref{PChannel} default constructor requires as minimum input a pointer to
an array of pointers to the parent and decay particles, and the number of
decay particles (default two):

\begin{itemize}
\item \classref{PChannel(PParticle **particles, int nt);}
\end{itemize}

Additional constructors are provided, adapted
to facilitate multi-hadron thermal decay modes of quasi-particle fireballs.

The decay models are not included in a single \classref{PChannel} by default, as
they are attached in a later step internally by using a dedicated interface in
order to allow the user to apply changes.  Therefore, \classref{PChannels} are
not working as stand-alone objects outside of a reaction.

\subsection{PReaction}\label{preaction}

\begin{itemize}
\item \classref{PReaction(PChannel **pchannel, char *file\_name, \\
    int n, int f0, int f1, int f2, int f3)}
\item \classref{PReaction(Double\_t momentum, \\
    char* beam, char* target, \\
    char* reaction, char* file\_name, \\
    Int\_t f0, Int\_t f1, Int\_t f2, Int\_t f3);}
\end{itemize}

A reaction is a complete physical process, consisting of one or several steps
(\classref{PChannels}). The \classref{PReaction} constructor requires as input
a \classref{PChannel}-type double pointer directed to an array of individual
\classref{PChannel}-type pointers, a character-string specifying a file name
including directory but without a suffix, and the number of constituent
channels (default is two).  Additional constructor arguments are flags of
integer type specifying output, decay-mode, and vertex-calculation options.
These are the following:

\begin{itemize}
\item[f0:] Output options for the ROOT file (default 0): 
  \begin{itemize}
  \item[0:] Only the tracked (i.e. stable) particles are stored in the ROOT file.
  \item[1:] All the particles are stored in the ROOT file, including the
    composite particles.  
  \end{itemize}

\item[f1:] Decay-mode options (default 0): obsolete
\item[f2:] Vertex-calculation options (default 0):
  \begin{itemize}
  \item[0:] This option is off (no vertex calculation).
  \item[1:] Production vertices are calculated for those particles that are
    written on file (depending on the output option). 
    The origin is considered to be the parent, or beam
    and target vertex. This is also the case for the products of the first
    channel. For particles produced in subsequent channels the production
    vertex is calculated by adding straight-line segments successively, each
    of length obtained as the product of the parent vector velocity times a
    lifetime randomly sampled from an exponential decay-time distribution.  An
    assumption of absence of any external magnetic fields is implicit.
  \end{itemize}
\item[f3:] ASCII output options (default 0):
  \begin{itemize}
  \item[0:] No ASCII output.
  \item[1:] ASCII output files, formatted for input to HGeant\footnote{HADES
    digitizer based on GEANT3~\cite{geant}.}.
    Irrespective of the output option, ASCII files contain only tracked
    particles. Invoked from the \classref{PDecayManager} (see below)
    class, a separate ASCII file is opened for each reaction channel
    processed.
  \item[2:] Common ASCII output file for all reaction channels processed by a
    \classref{PDecayManager}.
  \end{itemize}
\end{itemize}

\subsection{Decay manager}

\begin{figure}
\begin{center}
\resizebox{0.6\columnwidth}{!}{%
  \includegraphics{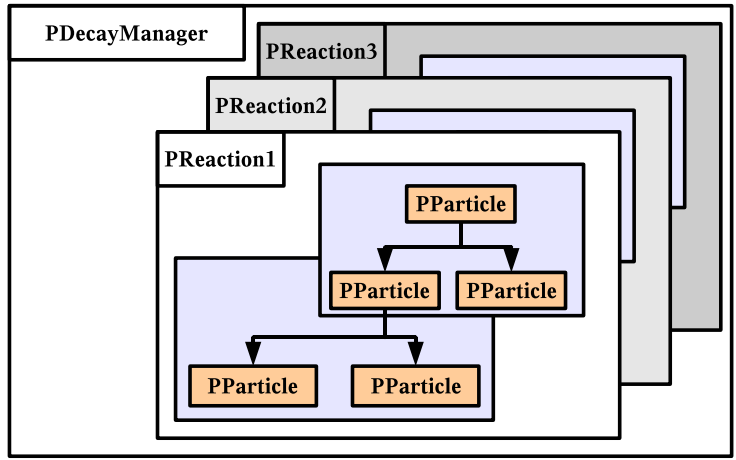}
}
  \caption{The decay manager: It combines several numbers of reactions in order
    to facilitate cocktail calculations. The relative weights of the reaction
    chains are laid down by the user or calculated by the tabulated branching
    ratio.}\label{fig:pdecaymanager}
\end{center}
\end{figure}

\subsubsection{Setup of the cocktail}

The \classref{PDecayManager} is a front end class for simulations that need
to cover a whole set of possible reaction chains (``cocktails''). It uses the
tabulated particle properties from the \classref{PDataBase}, including
(static) branching ratios.  Additional particles and decay branches may be
included via the member functions.

The standard way to set up a multi-step reaction is the following:

\subsubsection{Declaration of the PDecayManager}
{\footnotesize
\begin{verbatim}
PDecayChannel *c = new PDecayChannel;
PDecayManager *pdm = new PDecayManager;
pdm->SetVerbose(1);          // Print really useful info
\end{verbatim}}

\subsubsection{Preparation of the entrance channel}

First, the initial state is built (e.g. proton with 3.5 GeV kinetic energy on deuteron):

{\footnotesize
\begin{verbatim}
PParticle *p = new PParticle("p",3.5);  // proton beam
PParticle *d = new PParticle("d");      // deuteron target
PParticle *s = new PParticle(*p + *d);  // composite quasiparticle
\end{verbatim}}

\subsubsection{Preparation of the decay modes}

The user should assure that all the required channels are included in the
simulation. The most simple method is to use the static branching ratios as
defined in the data base and switch on any of the available channels, e.g.:

{\footnotesize
\begin{verbatim}
pdm->SetDefault("w");        // include omega decay modes
pdm->SetDefault("pi0");      // include pi0 decay modes
pdm->SetDefault("dilepton"); // e+e- production
\end{verbatim}}

\subsubsection{Declaration of the final state(s)}
Here, only one channel is used, but more channels can be added. If
'\classref{s}' is a predefined particle and has a list of decay modes
associated with it in \classref{PData}, '\classref{c}' can be omitted and the
predefined list can be used.  Otherwise the \classref{PDecayChannel} constructor (which is
a supporting class) may be utilized, as illustrated in the first two lines of
the sequence below. Once the final states have been specified, the reaction
can be initialized:

{\footnotesize
\begin{verbatim}
c = new PDecayChannel;    
c->AddChannel(0.1,"p","d","w");      // include decay modes
c->AddChannel(0.9,"p","p","pi0");    // include decay modes
pdm->InitReaction(s,c);              // initialize the reaction
\end{verbatim}}

The \classref{PDecayChannel} acts as list containing the default decay modes,
which is shown here with 2 decays channels with the relative weights (the
first number in the  \classref{AddChannel()}-method) selected such that the
$\omega$ is produced in 10\% of the events but the background from
$\pi^0$-production in 90\%. The last line actually combines the seed particle
(beam+target composite) with its default decay modes.

\subsubsection{Execution of the simulation}

{\footnotesize
\begin{verbatim}
pdm->loop(10000,0,"pdomega",1,0,0,0,0);   
\end{verbatim}}

The arguments are:

\begin{itemize}
\item[1.] Number of events: 10000. For reasons of normalization this number is
  the sum of event weights. The actual event number is returned by the loop
  function.
\item[2.] Weight flag: 0. If this is set, it acts as an additional
  normalization factor that adjusts the weight of the decay chain and all the
  product particles.  Otherwise, the number of events for one chain is
  calculated from the chain weight.
\item[3.] Reaction name: ``pdomega''. Used in setting up output file names.
\item[4.] Flags: f0, f1, f2, f3: These are the same as for \classref{PReaction}.
\item[5.] Random flag: if=0, process reactions sequentially, if=1, sample
  reactions in random order.
\end{itemize}

\subsubsection{Example: $\eta$ production}

In order to dicuss more applications for the distribution manager, the $\eta$
production in $pp$ collisions is shown.  Here, the $\eta$-meson can be
produced directly, or via the $N^*(1535)$ resonance. In order to avoid
implementing the definition of the $\eta$ decay many times in the macro code, the
\classref{PDecayManager} can be used, which finally constructs all reaction
chains which are possible:

{\footnotesize
\begin{verbatim}
// beam
PParticle p1("p",2.2);
PParticle p2("p");

PParticle q=p1+p2;
PDecayChannel * c = new PDecayChannel;
PDecayManager * pdm = new PDecayManager;

//primary meson production NONRESONANT
c->AddChannel(0.42,"p","p","eta");

//VIA N*(1535)
c->AddChannel(0.58,"p","NS11+");

//decay of the N*
PDecayChannel * nstar_decay = new PDecayChannel;
nstar_decay->AddChannel(1.0,"eta","p");
pdm->AddChannel("NS11+",nstar_decay);

//decay of the eta
PDecayChannel * eta_dalitz_decay = new PDecayChannel;
eta_dalitz_decay->AddChannel(1,"g","dilepton");
pdm->AddChannel("eta",eta_dalitz_decay);

//decay of the virtual photon:
PDecayChannel * eta_dilepton_decay = new PDecayChannel;
eta_dilepton_decay->AddChannel(1.0,"e+","e-");

pdm->AddChannel("dilepton",eta_dilepton_decay);
pdm->InitReaction(&q,c);

pdm->loop(100000,0,"eta_sample",1,0,0,1,1);
\end{verbatim}}

In this example, the decay modes are completely controlled by the user which
means that the $\eta$ Dalitz decay can be studied without background.

\subsection{Interface for bulk modifications}

Pluto allows for the seamless modification of the particle array (streaming
out after the defined reaction was performed) by external classes during the
loop execution. In this way, different event generators may interact with
Pluto (i.e. in order to execute the particle decay externally) and additional
particles can be added.  The latter method is very important for detector
studies, e.g. to embed single particle tracks into realistic background.  From
the software architecture point of view, the realization is done such that
objects inherited from the base class \classref{PBulkDecay} are added to the
\classref{PReaction}, which are executed after all decays (as defined in the
\classref{PReaction} constructor) have been finished. For external decays, at
the moment only one external event generator is considered which is
Pythia~\cite{pythia}.

\subsubsection{Embedded particles}

By using again the $\eta$ Dalitz example it is demonstrated how to add single
tracks in a reaction environment:

{\footnotesize
\begin{verbatim}
PReaction my_reaction(3.13,"p","p","p p eta [dilepton [e+ e-] g]",
    "eta_dalitz_embedded",1,0,0,0);

//Construct the bulk container:
PEmbeddedParticles * embedded = new PEmbeddedParticles();

//Add an e+ which we emit at a single point:
PParticle * e_plus = new PParticle("e+",1.,2.,3.);  
//Just add the particle to the container:
embedded->AddParticle(e_plus);

//We can also add a "white" dilepton, which we emit in a small cone:
PParticle * dilepton = new PParticle("dilepton");
embedded->AddParticle(dilepton);
embedded->SetSampling(0, 1.,   //pmin and pmax in lab frame 
                      TMath::Pi()/1000., //opening angle
                      TMath::Pi()/2.,    //Theta of pointing vect.
                      TMath::Pi()/2.,    //Phi of pointing vect.
                      0.2, 1.5);         //Mass sampling (optional)

//Add our container to the reaction:
my_reaction.AddBulk(embedded);

my_reaction.loop(100000);
\end{verbatim}}

where the class \classref{PEmbeddedParticles} is employed to carry one ore more
particle tracks which are (optionally) re-sampled during the event loop in order be
emitted in selected regions inside a detector setup. 

\subsubsection{Pluto bulk decay}

As it can be seen in the example above, the dilepton has been added in
addition to the $\eta$ Dalitz decay, but it was not decayed. Here (and in any
case where unstable particles are still un-decayed), the Pluto bulk decay class
can be used. The difference between the bulk decay and the normal treatment of
decays is that in the first case not a specific reaction chain is constructed,
but the particles are decayed using the mass-dependent branching ratio
(Eqn.~\ref{eqn:br}).  In order to demonstrate this feature, the decay of the
$N^*(1535)$ is used:

{\footnotesize
\begin{verbatim}
PReaction my_reaction(6,"p","p","p NS11+","n1535_sample_bulk",1,0,0,0);

PPlutoBulkDecay *pl = new PPlutoBulkDecay();
pl->SetRecursiveMode(1);  //Let also the products decay
pl->SetTauMax(0.001);     //maxTau in ns
my_reaction.AddBulk(pl);

my_reaction.loop(100000);
\end{verbatim}
}

\begin{figure}
  \begin{center}
    \resizebox{0.59\columnwidth}{!}{%
      \includegraphics{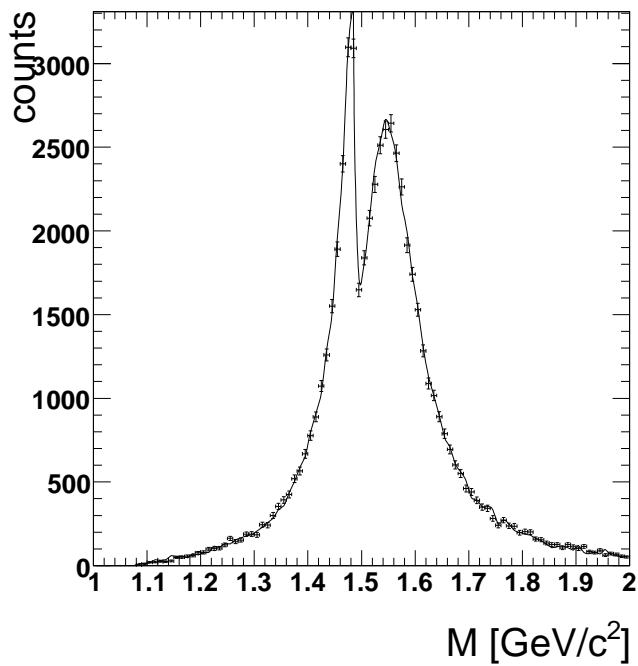}
    }
    \begin{minipage}[b]{0.40\columnwidth}
      \caption{The sampled $N^*(1535)$ mass shape using a selection on the final decay
        $N^*(1535) \to N + \pi$. The histograms have been obtained via two
        independent methods: the bulk decay (Solid line) and the  decay
        manager (data points).  Beside small phase space corrections the shape is equal to
        that one shown in Fig.~\protect\ref{fig:n1535_partial_contribution}.
        \newline ~}
      \label{fig:n1535_sample}       
    \end{minipage}
  \end{center}
\end{figure}

The decay is done recursively, until stable particles (defined such to have a
life-time larger then $\tau_{\rm max}$) are reached. This approach is
completely complementary to the \classref{PDecayManager} but should give the
same result beside the different normalization: The \classref{SetDefault}
method of the \classref{PDecayManager} is always based on the static branching
ratio (and in this way the relative weights of the included
\classref{PReactions} are calculated), whereas the mass-dependent branching
ratios are calculated by default to be fixed at the pole mass. If
Eqn.~\ref{eqn:scpw} is explicitly enabled (which is the case for the
$N^*(1535)$ e.g.), the result should be the same and provide an independent
cross-check of the framework functionality. For the decay of the $N^*(1535)$
in total 135 different possible reactions chains (combinations of possible
decays) are considered, and the sampling of 200.000 events including all decay
products takes usually 1 minute independent if the bulk decay or the decay
manager have been chosen.

In Fig.\ref{fig:n1535_sample} the obtained spectra for both methods are
compared. It can be seen that the distributions agree within the error bars, 
which gives confidence on the functionality of the framework. Moreover, they are
similar to the original calculation shown in
Fig.~\ref{fig:n1535_partial_contribution}.

\subsection{Distribution Manager}\label{pdistributionmanager}

The \classref{PDistributionManager} collects all information about the
included physics (single distributions and more complicated coupled channel
calculations) at one place and controls all objects which are offered to be
used in the \classref{PChannels}. The \classref{PChannels} instead do
not contain any calculation algorithm itself, but handle a list of
distributions matching the requirement given by each channel. The disabling of
distributions is done via unique identifier strings.  This ensures that the
production of events in reaction chains is done always consistent.

Therefore, the \classref{PDistributionManager} is a singleton which can be
obtained via a constructor method like:
{\footnotesize
\begin{verbatim}
PDistributionManager * dim = makeDistributionManager();
\end{verbatim}}
and the included physics can be listed with:
{\footnotesize
\begin{verbatim}
dim->Print();
\end{verbatim}}
The individual distributions are organized in groups, which are not expanded by default:
{\footnotesize
\begin{verbatim}
eta_physics      Physics about eta production, and decay: 5 enabled (from 5)
helicity_angles  Helicity angles of dileptons: 3 enabled (from 3)
resonances_pw    Partial waves of resonances: 4 enabled (from 4)
particle_models  Mass sampling of particles: 27 enabled (from 27)
decay_models     Phase space mass sampling & decay partial widths: 
                     178 enabled (from 178)
\end{verbatim}}
These groups can be completely enabled, disabled, or expanded via the following
commands:
{\footnotesize
\begin{verbatim}
makeDistributionManager()->Enable("helicity_angles");
makeDistributionManager()->Disable("helicity_angles");
makeDistributionManager()->ExpandGroup("helicity_angles");
\end{verbatim}}
The last command makes the distributions appearing in the \classref{Print()}-method,
alternatively \classref{Print("group\_id")} can be used:
{\footnotesize
\begin{verbatim}
...
    helicity_angles             Helicity angles of dileptons
[X] eta_dilepton_helicity       Helicity angle of the dilepton decay of eta
[X] etaprime_dilepton_helicity  Helicity angle of the dilepton decay of etaprime
[X] pi0_dilepton_helicity       Helicity angle of the dilepton decay of pi0
...
\end{verbatim}}
Single distributions can be enabled (or disabled) using their unique
identifier string as well:
{\footnotesize
\begin{verbatim}
makeDistributionManager()->Disable("eta_dilepton_helicity");
\end{verbatim}}
%
%
Another aspect is that distribution models might be strictly alternative and marked to
be not valid at the same time. This can be shown e.g. when printing the decay
models:
{\footnotesize
\begin{verbatim}
> makeDistributionManager()->Print("decay_models")
...
(X) w_picutoff_e-_e+     Dilepton direct decay with pion cutoff
    ( ) rho0_ee_e-_e+    Dilepton direct decay
...
\end{verbatim}}
Here, 2 independent models of the decay $\rho^0\to ee$ can be chosen, by
enabling one of them, the other will be disabled.

\subsection{Detector dependent aspects}\label{detector_filter}

In addition to the generation of events, Pluto includes filter and user
selection methods for rapid principle simulation studies adapted for
particular experimental conditions, detector setups and geometries.  A filter
object is instantiated by two arguments: a pointer to a reaction, and a
character string specifying explicitly the algebraic expression of the
condition that is to be satisfied. The \classref{PFilter} class implicitly
invokes the ROOT class \classref{TFormula} for the interpretation of the
character string, and for transposing it to a mathematical formula. A variety
of expressions are acceptable, inherited from the \classref{TFormula} class, including
trigonometric functions, exponentiation, and standard boolean logic working on
particle observables.

A general-purpose beam smearing model can be used to re-sample the beam
4-momentum event by event. Thus, simulations may take into account the
dispersion and (angular) resolution of the beam, adapted to the individual
experimental setup including the accelerator, and analyze the impact on
the signal resolution.

Finally, a general purpose file output interface makes it possible to define
output formats depending on the individual simulation framework of various
experiments. 

The points as mentioned above make it possible also for other experiments to
use Pluto and its included physics. On the other hand, it is very useful to
adapt in addition the build-in physics. This will be discussed in the next section.


\section{Customization}\label{customization}

Driven by the requirement of the on-going HADES analysis (and in parts also in
the context of simulations for the FAIR experiments) it is important that the
Pluto framework can be customized. Only by comparing the data with several
model assumptions conclusions can be drawn.  Therefore the Pluto framework was
extended\footnote{Included with v5.01 and v5.10.} such to make the
incorporation of new models possible which can be done be the user without
changing basic source code. This is achieved by a strictly object-oriented
design, in combination with the modular architecture.  The user is able to
interact with the Pluto-kernel in different ways, according to the level of
experience and needs:

\begin{enumerate}

\item The parameters of build-in distributions and models can be changed.

\item Build-in classes (serving as templates) can be utilized for reaction
  chains or decay channels not yet covered by build-in physics.
  
\item New classes may be created by the user and compiled during run-time of a
  macro. By implementing standardized interface methods, the control of decays
  or reaction chains can be taken over. These are named ``distribution
  models'' (base class \classref{PDistribution}).
  
\item New models may be created, which change in addition the partial widths
  and mass sampling of particles even if they are {\it not} directly used in
  the actual \classref{PReaction} setup. These are called ``channel models''
  (base class \classref{PChannelModel} inherited from
  \classref{PDistribution}).

\end{enumerate}

\subsection{Changing the build-in implementation parameters}
\label{change_params}

The way to change parameters can be demonstrated using e.g. the $\eta
\to\pi^+\pi^-\pi^0$ matrix element. Recent results~\cite{cbarrel2} differ from
the build-in distribution, and new generation experiments are going to
re-measure the Dalitz slopes with a high precision~\cite{wasa_kupsc}. These
slopes can be changed by obtaining first the object from the
\classref{PDistributionManager} by the known identifier:

{\footnotesize
\begin{verbatim}
PDistributionManager *pdist = makeDistributionManager();
PDalitzDistribution  *eta_pion_decay = (PDalitzDistribution *)
    pdist->GetDistribution("eta_hadronic_decay");
\end{verbatim}}

Consequently, the local methods of the given distribution class can be used:

{\footnotesize
\begin{verbatim}
eta_pion_decay->SetSlopes(1.22,0.22); //new result
\end{verbatim}}

\subsection{Re-use build-in distributions}

\begin{figure}
\begin{center}
\resizebox{0.9\columnwidth}{!}{%
  \includegraphics{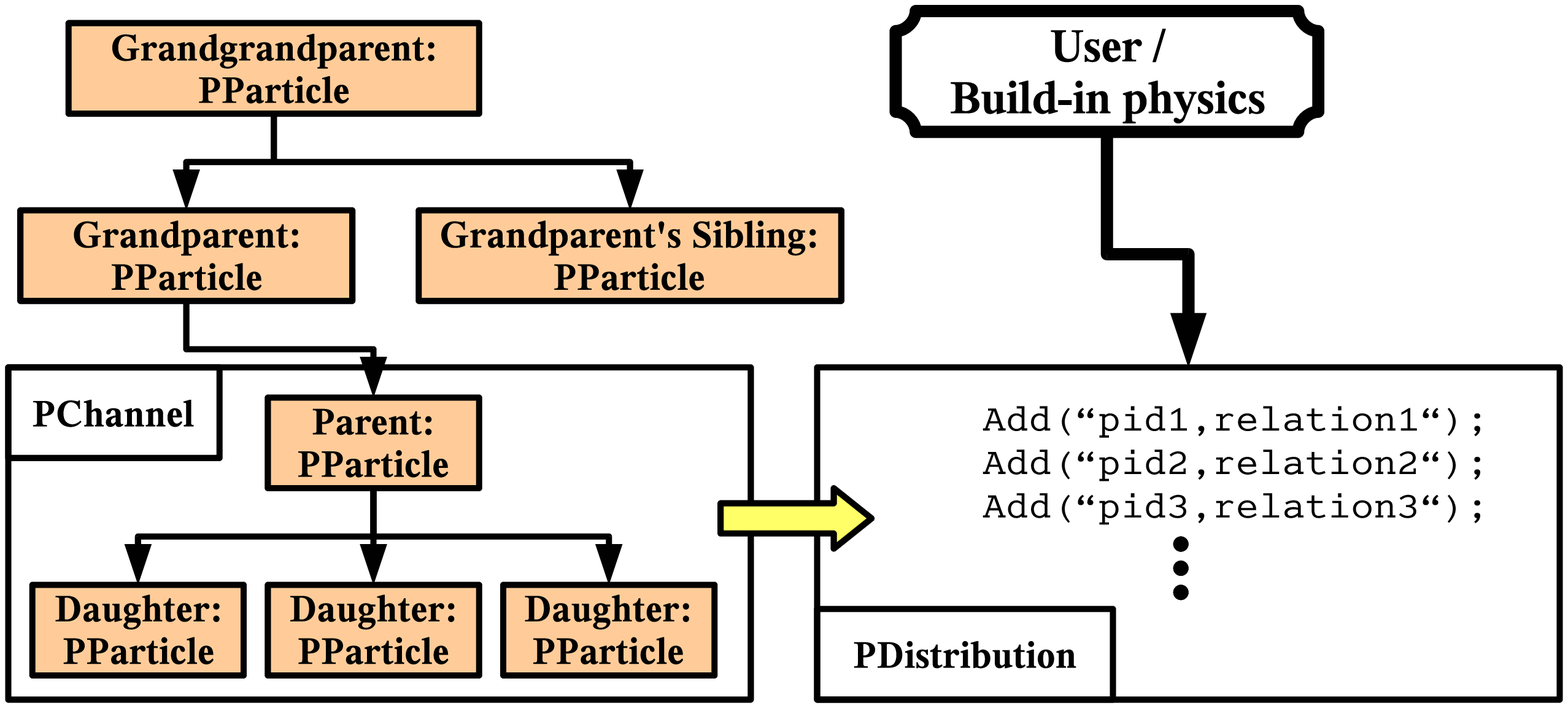}
}
\caption{Basic idea of the connection between a single \classref{PChannel} and the
  \classref{PDistributions} (the base class for all distributions): The user adds a
  list of tracks as a template. Each track is identified by a particle id
    and a relationship (e.g. daughter, parent, (grand)grandparent). During
    runtime, this template is being compared to the reaction chain as given by
    the \classref{PChannel}, and the \classref{PDistribution} is attached on
    success.}\label{fig:pdistribution_attach}
\end{center}
\end{figure}

The basic concept of the \classref{PDistribution} (and all distributions
inherited from this base class) is that they can be implemented many times for
different reaction chains, including the (grand)grandparent, and their
siblings in addition. For of these each implementation procedure the user has
to perform the following steps:

\begin{enumerate}

\item The object has to be created with the \classref{new} operator, among with a title
  and identifier
  (the latter one must be unique).

\item The user defines a decay chain template with the \classref{Add()}-method. The right half of
  Fig.~\ref{fig:pdistribution_attach} gives an overview of this principle.

\item All local parameters which are needed for the individual algorithm have
  to be initialized correctly according to the physics case. These methods
  depend on the class which has been used in this context.

\item The new object has to be added to the \classref{PDistributionManager}.

\end{enumerate}

\subsection{Angular distributions}\label{angularcase}

In order to show how a distribution class is implemented, angular
distributions are discussed, as they play always an important role in the
production and decay of mesons and resonances. They could be sensitive to the
production mechanism (exchange of bosons), and might effect in combination
with the detectors acceptance the integrated yield for individual channels.
Therefore, attention should be paid on the sensitivity of specific angular
distribution cases. Here, it is used as example to demonstrate the usage of a
build-in distribution class (\classref{PAngularDistribution}), which is
able to handle many general purpose cases: Polar angle distributions in the production of
particles and polar alignment, as needed e.g.  for the $\eta$ physics.

The first step it to instantiate the object:

{\footnotesize
\begin{verbatim}
PAngularDistribution * my_distribution = new
    PAngularDistribution("meson_distribution", 
    "Polar angle of my meson");
\end{verbatim}}
  
  Here, the first parameter is the unique identifier, and the second is a
  title. Now, the decay chain template has to be created, as depicted in
  Fig.~\ref{fig:pdistribution_attach}, which means that at least the parent
  and the daughters must be defined to be unambiguous.
  
  On the other hand in the case of - let us say the emission of a $\pi^0$ in $pp$
  collisions - this information is not enough, because the angular
  distribution algorithm has to know {\it which} of the particles in the three
  body final state has to be sampled. In order to deal with such ambiguities
  in the internal calculations the decay chain template can be extended
  with ``private'' option flags which can be read by the distribution objects
  during the initialization procedure. Here, the angular distribution class is
  able to handle different cases explicitly, by choosing the primary particle
  by such a flag, and its reference frame.
  
  For a first simple example, it should be assumed that the primary particle
  (the particle for which the polar angle has to be sampled) is a $\pi^0$
  meson, and the reference for the angle is the c.m.  frame:

{\footnotesize
\begin{verbatim}
my_distribution->Add("pi0,daughter,primary");
my_distribution->Add("p,daughter");
my_distribution->Add("p,daughter");
my_distribution->Add("q,parent,reference");
\end{verbatim}}

In addition to the pid names (as listed in the data base), the distribution interface
offers in addition helpful identifiers:

\begin{description}

\item[``q'':] Any composite particle
\item[``?'':] A wildcard for any track
\item[``*'':] Any number of additional tracks
\item[``N'':] Any nucleon

\end{description}

Each particle is finally combined with one of the possible relationships:
``parent'', ``(grand)grandparent'', or siblings of such relatives (like
``parent,sibling'').

In total, the given template reads like this: The distribution, which has been
created in our example will be attached to any
\classref{PChannel}, which has a composite particle as a parent and 3 selected
decay products (one
$\pi^0$ and 2 Protons).

The next step is to define the angular distribution function. The class
\classref{PAngularDistribution} uses TF1/TF2 ROOT objects for this purpose.
The latter one is used for mass-dependent parameterizations like in
Eqn.~(\ref{c2_param}). A simpler application would be to define a distribution
only depending
on $x=\cos^2\theta$:

{\footnotesize
\begin{verbatim}
TF1 *angles=new TF1("angles","(x*x)/2",-1,1);
my_distribution->SetAngleFunction(angles);
\end{verbatim}}
Finally, the new distribution has to be added to the distribution manager:
{\footnotesize
\begin{verbatim}
makeDistributionManager()->Add(my_distribution);
\end{verbatim}}
  
\begin{figure}
\resizebox{0.07\columnwidth}{!}{~}
\resizebox{0.4\columnwidth}{!}{%
  \includegraphics{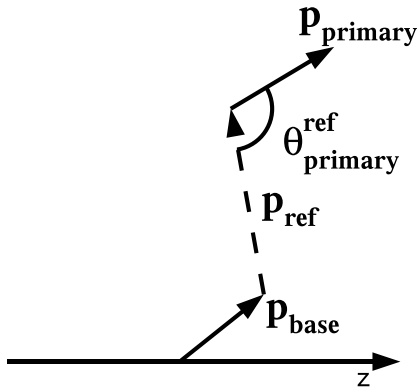}
}\resizebox{0.07\columnwidth}{!}{~}
\begin{minipage}[b]{0.45\columnwidth}
\caption{
Definition of the polar angle in the \classref{PAngularDistribution} class: The
  polar angle of the particle ``primary'' is defined with respect to the
  momentum of the particle ``reference'' (The parent by default). This
  momentum, however, is defined in the ``base'' frame as a reference (c.m. by
  default).\newline ~} 
\label{fig:angles_definition}
\end{minipage}
\end{figure}

  The discussed class \classref{PAngularDistribution} allows also for the sampling
  of more sophisticated angular correlations. Resonances like $N^*(1440)$
  should decay into $p\pi$ with dedicated partial waves, hence the
  dashed line in Fig.~\ref{fig:angles_definition} is the $N^*$-momentum in the
  c.m. frame. In this context, the configuration would have to be defined
  like:

{\footnotesize
\begin{verbatim}
my_distribution->Add("pi0,daughter,primary");
my_distribution->Add("p,daughter");
my_distribution->Add("NP11+,parent,reference");
my_distribution->Add("q,grandparent,base_reference");
\end{verbatim}}
  
  By using exactly such a method the $\Delta \to N \pi$ decay angular
  distribution (discussed in Sec.~\ref{pw_delta}) was implemented.

\subsection{Changes in the data base}\label{database}

Pluto uses a multi-purpose data base with fast lookup-keys to store all
information which are needed to read the particle properties as well as for the
handling of the models. Internally, these data base is filled in 2 steps:
First, when user-interface classes like \classref{PChannel} or
\classref{PParticle} are created the first time, the interface singleton
\classref{makeStaticData()} fills the particle properties from a fixed table.
Second, if a \classref{PReaction} or \classref{PDecayManager} is instantiated
the \classref{PDistributionManager} is filled and linked to the data base.
Tool classes are facilitated to create dedicated models for each decay using the
particle properties like lepton number or width.
 
This has two consequences for the user: particle and decay properties should be
changed before the first creation of a particle takes place, and models should be chosen
using the \classref{PDistributionManager}-\classref{Enable()} or
-\classref{Disable()} method before creating the reaction.
 
The direct write access to the data base is possible but recommended only
for experts, however for the normal usage a various number of methods are
available by using the \classref{makeStaticData()} interface. An example of
defining a new decay like $A \to b+c$ is depicted below:

{\footnotesize
\begin{verbatim}
makeStaticData()->AddParticle(-1,"A", 1.2);
makeStaticData()->SetParticleTotalWidth("A",0.3);
makeStaticData()->AddParticle(-1,"b", 0.5);
makeStaticData()->AddParticle(-1,"c", 0.3);
makeStaticData()->AddDecay(-1,"A -> b + c", "A", "b,c", 1.);
\end{verbatim}}
  where the ``$-1$'' indicates that Pluto should assign the next available pid and
  decay index number. The numbers used in the methods are the mass (for
  particles) and the branching ratio (for decays). The latter one are
  re-normalized such that the total sum is always equal to one.

\subsection{Adding new models}\label{models}

A complete description how to implement new distribution and channel models
would go beyond the scope of this report.  However, a short sketch of the
technical realization in the framework is presented which is useful in the
context of the discussion in Sec.~\ref{physics} and~\ref{angular}. The idea is
that during the event loop the methods of the attached models in each
\classref{PChannel} are called which allow each distribution object to update
the particle properties without any restriction, these are:

\begin{figure}
\begin{center}
\resizebox{0.43\columnwidth}{!}{%
  \includegraphics{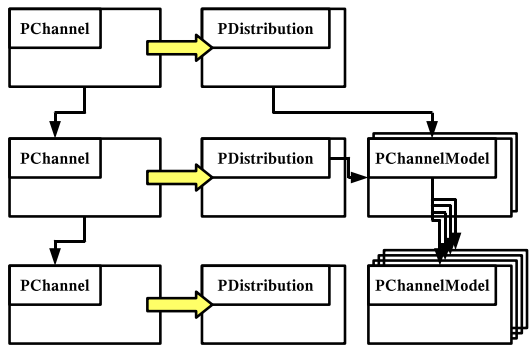}
}
\resizebox{0.04\columnwidth}{!}{%
  ~
}
\begin{minipage}[b]{0.50\columnwidth}
\caption{
  Technical representation of the coupled channel calculations done inside the
  Pluto-framework: In the \classref{PChannel}-chain, the
  \classref{PDistribution} objects are called via defined interface methods in
  order to sample e.g. masses and momenta. As these involve all decays (see
  e.g. Eqn.~\protect\ref{eqn:width_sum}) hidden channels are called even if
  they do not participate in the chosen decay chain.
}\label{fig:coupled_channel}
\end{minipage}
\end{center}
\end{figure}

\begin{itemize}

\item \classref{Init()}: Called only once before the event loop is started to allow
  the distribution to obtain the particle objects via the private identifiers
  as discussed in Sec.~\ref{angularcase}.
  
\item \classref{Prepare()}: Initial action in each event of the loop. Used
  e.g. for the beam smearing model shortly introduced in
  Sec.~\ref{detector_filter}.

\item \classref{SampleMass()}: Change the mass of the particles. Used e.g. for
  the resonance sampling.

\item \classref{SampleMomentum()}: Used by the Genbod-model and Fermi momentum
  sampling.

\item \classref{SampleAngle()}: Used by the models to sample polar angle
  distributions.
  
\item \classref{IsValid()}: Used by rejection models, e.g. the $pp$ polar
  alignment in the $pp\to pN^* \to pp \eta$ reaction, because in such a case
  it is not possible to obtain an analytical function in the c.m. frame.

\item \classref{CheckAbort()}: Can be used by the models to force the \classref{PReaction}
  to abort the complete chain and start from the decay of the first channel.

\item \classref{Finalize()}: Final actions.

\item \classref{GetLocalWeight()}: Update the particle weight of the daughters. 

\end{itemize}

For the coupled channel models, the situation is more complex, since they
should be able to be called without existing particle objects.
Fig.~\ref{fig:coupled_channel} sketches this idea: A very large number of
hidden channels might be involved which in particular the case in the sampling of
higher resonance masses. The reason is that as explained in
Sec.~\ref{m3_model} decay sampling calculations involve the mass shapes and
vice versa. 

In order to consider this, coupled channel classes add
some more methods. These methods are {\it not} directly called by the 
\classref{PChannel} as depicted in Fig.~\ref{fig:coupled_channel}, they use a 
\classref{PDistribution} (which is bound always to a decay) as a doorway to
the coupled-channel models. 
We use 2 variants of channel models, namely:

\begin{enumerate}
\item Models used for particle mass shapes for the calculation of $g(m)$ (or
  $g^{\rm k}(m)$), and:
\item For decays in order to calculate $G^{\rm k}(m_1, m_2, \dots )$. These
  include the determination of the partial decay widths.
\end{enumerate}

The latter one may work in addition as the interface to the
\classref{PChannel}.  Both implementations use one or more of the
following methods:

\begin{itemize}

\item \classref{GetWeight(Double\_t *mass, Int\_t *decay\_index)}:
  Calculating the weight, either $g(m)$ or $G(m_1, m_2, \dots )$. If the
  \classref{decay\_index} array is defined, the functions $g(m)$ and $G(m_1, m_2, \dots
  )$ have to be replaced by $g^{\rm k}(m)$ and $G^{\rm k}(m_1, m_2, \dots
  )$.

\item \classref{GetAmplitude(Double\_t *mass, Int\_t *decay\_index)}: Returns
  the complex amplitude as used in the context of Sec.~\ref{complex}.

\item \classref{GetWidth(Double\_t *mass, Int\_t *decay\_index)}: Calculates
  the mass-dependent width as discussed in Eqns.~(\ref{eqn:m1_width},\ref{eqn:m2_width}).
  
\item \classref{SampleMass(Double\_t *mass, Int\_t *decay\_index)}: Some as
  above but without particle objects.
      
\item \classref{GetBR(Double\_t mass, Double\_t *br)} Calculates
  mass-dependent branching ratios.

\end{itemize}

As a final summary of the implemented framework, let us consider what will be
done if a reaction has been defined by the user, like
{\footnotesize
\begin{verbatim}
PReaction my_reaction(3.13,"p","p","p NS11+ [p eta [dilepton [e+ e-] g] ]",
    "nstar",1);
\end{verbatim}}
  This constructor first calls the particle data filler in order to set up the
  data base. The particle decays are normalized such that the sum of all
  branching ratios will be equal to one.  The ``+'' operator involved in the
  beam-target interaction adds the composite to the data base (which is
  further treated a normal particle). All involved \classref{PChannels} are
  constructed by the \classref{PReaction} among with their
  \classref{PParticles}. In a second step, the \classref{PDistributionManager}
  is instantiated and filled with all models using an included logic on e.g.
  particle species and width. Each of these coupled-channel models is
  linked to the data base.
  Moreover, all distribution models are attached to their corresponding
  \classref{PChannels} (if existing) and initialized, thus they have access to
  the particle objects.

By starting the loop:
{\footnotesize
\begin{verbatim}
my_reaction.Loop(10000)
\end{verbatim}}
  the distribution interface methods are called in each step as defined above.
  At least one model per channel is required as a doorway to do the mass
  sampling, in our example it is the decay in one stable and one unstable particle as
  outlined in Eqn.~\ref{eqn:m2_mass}. Hidden model are called to sample the
  $N^*(1535)$ mass shape taking all its decay modes into account to get each individual
  partial decay width. This involves moreover the mass shape of other
  resonances (e.g. the $N^*(1440)$).  
  
  In the case that a selected chain is calculated, as in our example, mass
  sampling is done in each step using the partial decay width rather then the
  total, which is automatically assigned by the framework.
  
  After the decay products of the first \classref{PChannel} have been sampled
  the second \classref{PChannel} is called. On one hand this is a simple decay of a
  resonance into two stable products. On the other hand,
  the angles as defined in Figs.~\ref{fig:meson_polar_angles}
  and~\ref{fig:angles_definition} are determined by the class
  \classref{PAngularDistribution} and used to sample the angular distribution
  function. On failure of the rejection method the complete chain is
  re-sampled thus avoiding any distortion of the $N^*(1535)$ mass shape.

  The third decay involves the mass sampling of the $\eta$ Dalitz decay,
  discussed in Sec.~\ref{etadalitz}. The sampled virtual photon is been
  decayed in a forth step into leptons with obviously fixed masses.

  This demonstrates that many cases are handled in
  background without any required user interaction.



\section{Summary}\label{summary}

In summary, we presented the Pluto framework, originally intended for studies
with the HADES detector. It is based on C++  and ROOT and has a very user-friendly
interface, starting from a few lines of code the event production for selected
channels can be initialized.

The standardized interface between the event loop and the models allows for the
implementation of customized model classes. Several interface methods may be used,
for mass and momentum sampling up the handling of complex amplitudes in
hidden coupled-channel calculations. This enables the calculation of spectral
functions from first principles for hadronic resonances with multiple decay
modes. This capability, together with a number of theoretical and empirical
hadronic-interaction models implemented in the code, provide tools for
realistic simulations of elementary hadronic interactions, such as resonance
excitation and decay, elastic proton-proton scattering, Dalitz decays, and
direct dilepton decays of vector mesons.

Several empirical parameterizations on angular distributions and momentum
sampling have been included. Moreover, a thermal model has been developed.  The
latter one handles multi-hadron decays of hot fireballs, and provide tools for
studies of thermally produced hadrons and the distributions of their
observables, comparison studies and the subtraction of trivial sources.  The
addition of a decay manager interface trivializes the setting up and execution
of multi-channel (``cocktail'') simulations in elementary collisions.

The package presented has been used to perform simulations for various HADES
experiment proposals, for the comparison of the HADES C+C heavy ion data and
for the $pp$ model-dependent acceptance corrections. Moreover, it has been
utilized in the simulation for the coming CBM experiment at FAIR.

The versatility and re-usability of the code, as demonstrated by a number of
user interface classes and simulation macros distributed with the package,
allow for rapid principle simulation studies adapted for particular questions.
Moreover, it is fully object-oriented, thus any specific process model can be
exchanged by the user and defined interface methods allow for the interaction
with the framework.

Therefore, it is open for future developments, as this is needed for the
upcoming HADES experiments and simulations for the new FAIR facility at GSI.

\section{Acknowledgments}

The original author of this package (M.~Kagarlis) likes to thank Christoph
Ernst for his input regarding theoretical models and their implementation, and
Rene Brun for his frequent assistance with ROOT-related questions, at the
early stages of this project.  Moreover, he is grateful to Richard Arndt,
whose seminal work on phase-shift analyses of $NN$ and $\pi N$ elastic
scattering data over the years is the standard in the literature, who has
kindly supplied an algorithm from his code (SAID) that is implemented for $pp$
elastic scattering angular distributions. He finally wishes to
express his gratitude to Bengt Friman, whose input has been invaluable in
developing the models that have been implemented in the framework for the
calculation of hadronic-resonance widths and spectral functions.

The first author of this report thanks C.~Sturm and M.~Bleicher for the
fruitful discussion on the $\rho^0$ mass shape, and E.L. Bratkovskaya for 
proofreading parts of the manuscript. Thanks also to A.~Kupzc for the
recent hints on the $pp$ elastic channel.

This work was supported by the GSI (in particular the International Student
Summer School of GSI), the BMBF and the EU under the contract CNI
(construction) ``DIRAC-PHASE-1'' (515876).

\end{document}